\documentclass[]{spie}  %>>> use for US letter paper
\usepackage[utf8]{inputenc}
\usepackage{multirow}

\usepackage{amsmath,amsfonts,amssymb}
\usepackage{graphicx}
\usepackage[colorlinks=true, allcolors=blue]{hyperref}

\title{Beyond CCDs: Characterization of sCMOS detectors for optical astronomy}

\author[a]{Aditya Khandelwal}
\author[b]{Sarik Jeram}
\author[b]{Ryan Dungee}
\author[b]{Albert W.K. Lau}
\author[a]{Allison Lau}
\author[a]{Ethen Sun}
\author[a, b]{Phil Van-Lane}
\author[b]{Shaojie Chen}
\author[a]{Aaron Tohuvavohu}
\author[a, b]{Ting S. Li}
\affil[a]{David A. Dunlap Department of Astronomy and Astrophysics, University of Toronto, 50 St. George Street, Toronto, ON, Canada M5S 3H4}
\affil[b]{Dunlap Institute for Astronomy and Astrophysics, University of Toronto, 50 St. George Street, Toronto, ON, Canada M5S 3H4}

\authorinfo{Further author information: \\
Aditya Khandelwal: adi.khandelwal@mail.utoronto.ca}

\begin{document} 
\maketitle

\begin{abstract}
Modern scientific complementary metal-oxide semiconductor (sCMOS) detectors provide a highly competitive alternative to charge-coupled devices (CCDs), the latter of which have historically been dominant in optical imaging. sCMOS boast comparable performances to CCDs with faster frame rates, lower read noise, and a higher dynamic range. Furthermore, their lower production costs are shifting the industry to abandon CCD support and production in favour of CMOS, making their characterization urgent. In this work, we characterized a variety of high-end commercially available sCMOS detectors to gauge the state of this technology in the context of applications in optical astronomy. We evaluated a range of sCMOS detectors, including larger pixel models such as the Teledyne Prime 95B and the Andor Sona-11, which are similar to CCDs in pixel size and suitable for wide-field astronomy. Additionally, we assessed smaller pixel detectors like the Ximea xiJ and Andor Sona-6, which are better suited for deep-sky imaging. Furthermore, high-sensitivity quantitative sCMOS detectors such as the Hamamatsu Orca-Quest C15550-20UP, capable of resolving individual photoelectrons, were also tested. In-lab testing showed low levels of dark current, read noise, faulty pixels, and fixed pattern noise, as well as linearity levels above $98\%$ across all detectors. The Orca-Quest had particularly low noise levels with a dark current of $0.0067 \pm 0.0003$ e$^-$/s (at $-20^\circ$C with air cooling) and a read noise of $0.37 \pm 0.09$ e$^-$ using its standard readout mode. Our tests revealed that the latest generation of sCMOS detectors excels in optical imaging performance, offering a more accessible alternative to CCDs for future optical astronomy instruments.
\end{abstract}

% Include a list of keywords after the abstract 
\keywords{sCMOS, CCD, optical astronomy, detectors, photon counting, sensor noise}

\section{INTRODUCTION} \label{sec: intro}

For most of human history, recording the night sky relied on drawings and written descriptions, which were often inaccurate and limited by the human eye's arcminute resolution. The invention of the Galilean telescope in 1609 marked a major technological shift\cite{telescopes}, allowing astronomers to see further and with greater detail, though they still had to manually document their observations. The advent of astronomical imaging in 1840, using photographic plates, revolutionized observational astronomy by enabling precise, permanent records of celestial objects\cite{Astrophotography1}. This advancement allowed astronomers to accurately measure the time evolution and long-term motion of celestial bodies, transforming the field.

In 1975, the introduction of silicon-based charge-coupled devices (CCDs) revolutionized astronomy again by using the photovoltaic effect to capture digital photographs of the night sky\cite{CCDs, sensor_history}. Combined with computers, these digital images allowed for much higher precision and accuracy in measuring light from celestial bodies, significantly expanding the scope of astronomical research. CCDs offered numerous advantages over photographic plates, including vastly improved sensitivity, quantum efficiency, dynamic range, responsiveness to a broader spectrum of light, a linear response to intensity and exposure time, and the ability to produce digital images that could be immediately processed and corrected\cite{sensor_history}.

Complementary metal-oxide semiconductors (CMOS) are another silicon-based technology that uses the photovoltaic effect to capture images\cite{CMOS}. Unlike CCDs, which use a shift register for pixel readout, each pixel in a CMOS chip has its own readout electronics, effectively making CMOS a collection of individual detectors. Historically, CMOS detectors underperformed compared to CCDs, producing lower quality images with lower sensitivities, non-linear responses, and high fixed pattern noise (FPN), leading to CCDs dominating scientific use\cite{CMOS_performance}.

However, CMOS detectors offered advantages such as faster frame rates, low read noise, lower power consumption, and resistance to visual artefacts like blooming and smearing\cite{CMOS_performance}. Their cost efficiency and low power consumption made them popular in consumer electronics, leading to increased development and production improvements. By 2009, advancements in CMOS technology led to the development of scientific-grade CMOS detectors with performance on par with CCDs, while retaining their inherent benefits\cite{sCMOS}.

Due to the large consumer market for CMOS, CCD manufacturing is becoming increasingly limited, making them harder and more expensive to acquire. Both Sony and ON Semiconductors have ceased CCD production, with ON Semiconductors' last sales in 2020\cite{onsemi} and Sony planning to end all CCD support by 2026\cite{sony}. Consequently, it is crucial to characterize high-end commercially available sCMOS detectors and evaluate if their performance is competitive enough to prompt a shift in scientific astronomy imaging.

To assess the suitability of sCMOS technology for optical astronomy applications, it is essential to comprehensively evaluate the performance of sCMOS models from various manufacturers. These models, while not specifically tailored for astronomy, must have varying pixel and array sizes that can be advantageous for different scientific purposes within the field. To comprehensively characterize a detector, we must measure its electronic and thermal noise contributions, given by the read noise and dark current respectively, the linearity of its signal response, and its quantum efficiency within the visible spectrum.

In this paper, we characterize a range of high-end commercially available backside-illuminated sCMOS detectors from different vendors. These detectors were selected based on pixel size, price point, manufacturer-reported performance, and demo availability. We assessed properties such as read noise, dark current, linearity, and quantum efficiency.

In Section \ref{sec: detectors}, we offer a brief overview of all the detectors characterized in this work, along with some manufacturer-reported performance metrics. Section \ref{sec: method} covers the data collection and analysis methods used for the various tests. In Section \ref{sec: results}, we summarize the characterization results for all detectors, with a detailed discussion on the Orca-Quest results. Finally, Appendix \ref{sec: appendix} provides additional details on the characterization results of the other detectors.

\section{DETECTORS} \label{sec: detectors}

To get a comprehensive understanding of the current state of sCMOS technology, we characterized five different detectors in this work, categorizing them into three broad groups: large pixel detectors suited for large sky surveys, small pixel detectors ideal for deep field imaging, and high-sensitivity detectors with photon number resolving capabilities. When possible, we also characterized multiple readout modes provided by some detectors.

The Teledyne Prime 95B (25mm), hereafter referred to simply as Prime 95B, uses a 2.6-megapixel GPixel GSense 400 BSI sensor with $11 \times 11$ $\mu$m pixels\cite{teledyne_tech_note}. The Prime 95B has three readout modes that were characterized: balanced, full well, and sensitivity. The full well mode utilizes the detector's maximum full well capacity, the sensitivity mode maximizes sensitivity, while the balanced mode uses an optimized midpoint between sensitivity and full well\cite{teledyne_tech_note}. All three modes are 12-bit, with a saturation limit of 4096 ADU. The Prime 95B boasts a peak quantum efficiency $>95\%$\cite{teledyne_tech_note}.

The Andor Sona-11 (32 mm), simply called Sona-11, uses a 4.2-megapixel Sona 4.2B-11 sensor with $11 \times 11$ $\mu$m pixels\cite{andor_tech_note}. Only the 16-bit high dynamic range mode of the Sona-11 was characterized with a saturation limit of 65536 ADU. The Sona-11 boasts a peak quantum efficiency of $95\%$\cite{andor_tech_note}.

The Andor Sona-6 Extreme, simply called Sona-6, uses a 4.2-megapixel Sona 4.2B-6 sensor with $6.5 \times 6.5$ $\mu$m pixels\cite{andor_tech_note}. Only the 12-bit low noise mode of the Sona-6 was characterized with a saturation limit of 4096 ADU. The Sona-6 boasts a peak quantum efficiency of $95\%$\cite{andor_tech_note}.

The Ximea xiJ MJ042MR-GP-P6-BSI, simply called Ximea XiJ, uses a 4.2-megapixel GPixel GSense 2020 BSI sensor with $6.5 \times 6.5$ $\mu$m pixels\cite{ximea_tech_note}. Two readout modes of the Ximea xiJ were characterized: standard and correlated multiple sampling. Both modes are 12-bit, with a saturation limit of 4096 ADU. The Ximea xiJ boasts a peak quantum efficiency of $91\%$\cite{ximea_tech_note}.

The Hamamatsu Orca-Quest CP15550-20UP, simply called Orca-Quest, is advertised as being a quantitative CMOS detector with extremely low noise levels and photoelectron counting capabilities. It features a custom 9.4-megapixel sensor with $4.6 \times 4.6$ $\mu$m pixels\cite{hama_tech_note}. The Orca-Quest has two scan modes that were characterized: standard and ultra-quiet. The ultra-quiet mode has a much lower frame rate at 5 frames per second (fps) compared to the standard mode's 120 fps, which allows for much lower read noise\cite{hama_tech_note}. Also characterized was the `photon number resolving' readout mode which claims to report the integer number of incident photoelectrons based on a proprietary calibrated algorithm using the ultra-quiet scan\cite{hama_tech_note}. The Orca-Quest has a detector-imposed temperature lock at $-20^\circ$C when air-cooled. The standard and ultra-quiet modes are 16-bit, with a saturation limit of 65536 ADU while the photon number resolving mode has a saturation limit of only 200 ADU. The Orca-Quest boasts a peak quantum efficiency of $85\%$\cite{hama_tech_note}.

The Orca-Quest was purchased for more detailed testing while the other detectors were only tested within their time-constrained demo periods.

 The full well capacity, pixel size, array size, and reported global performances of each detector in their standard readout mode are summarized in Table \ref{tab: detectors}.

\begin{table}[h!]
    \caption{Summary of the detectors selected for characterization and their manufacturer-reported global performances in their standard readout mode\cite{teledyne_tech_note, andor_tech_note, ximea_tech_note, hama_tech_note}. \underline{Note:} For the Prime 95B the `standard' mode is the sensitivity mode. For the Sona-11 and the Sona-6, the standard mode simply refers to the singular mode tested for both detectors. The Sona-6 also features a 16-bit high dynamic range mode with a 42000 e$^-$ full well capacity that was not characterized due to time constraints.}
    \label{tab: detectors}

    \begin{center} 
    \begin{tabular}{c|cccccc}
    
    Detector & 
    \begin{tabular}[c]{@{}c@{}}Full well capacity \\{[}e$^-${]}\end{tabular} & 
    \begin{tabular}[c]{@{}c@{}}Pixel size\\ {[}$\mu$m{]}\end{tabular} & 
    \begin{tabular}[c]{@{}c@{}}Array size\\ {[}pixels{]}\end{tabular} & 
    \begin{tabular}[c]{@{}c@{}}Read noise\\ {[}e$^-${]}\end{tabular} & 
    \begin{tabular}[c]{@{}c@{}}Dark current\\ (air cooled)\\ {[}e$^-$/s{]}\end{tabular} & 
    Linearity \\ 
    \hline
    \hline
    
    {\begin{tabular}[c]{@{}c@{}}Teledyne Prime\\95B (25 mm)\end{tabular}} 
                           & 80000 & $11 \times 11$ & $1608 \times 1608$
                           & 1.6 & {\begin{tabular}[c]{@{}c@{}}0.55 \\ $@ -20^\circ$C\end{tabular}} & $>99.5\%$ \\
                           \hline
    
    \begin{tabular}[c]{@{}c@{}}Andor Sona-11\\ (32 mm)\end{tabular}
                           & 85000 & $11 \times 11$ & $2048 \times 2048$
                           & 1.6 & {\begin{tabular}[c]{@{}c@{}}0.7 \\ $@ -25^\circ$C\end{tabular}} & $>99.7\%$ \\
                           \hline
    
    \begin{tabular}[c]{@{}c@{}}Andor Sona-6\\ Extreme\end{tabular}
                           & 1100 & $6.5 \times 6.5$ & $2048 \times 2046$
                           & 1.0 & {\begin{tabular}[c]{@{}c@{}}0.15 \\ $@ -25^\circ$C\end{tabular}} & $>99.7\%$ \\
                           \hline

    {\begin{tabular}[c]{@{}c@{}}Ximea xiJ \\ MJ042MR-\\ GP-P6-BSI\end{tabular}}
                           & 54000 & $6.5 \times 6.5$ & $2048 \times 2048$
                           & 1.8 & {\begin{tabular}[c]{@{}c@{}}0.2 \\ $@ -20^\circ$C\end{tabular}} & $>96.8\%$ \\
                           \hline

    {\begin{tabular}[c]{@{}c@{}}Hamamatsu\\ Orca-Quest\\ C15550-20UP\end{tabular}}  
                           & 7000 & $4.6 \times 4.6$ & $4096 \times 2304$
                           & 0.43 & {\begin{tabular}[c]{@{}c@{}}0.016 \\ $@ -20^\circ$C\end{tabular}} & $>99.5\%$ \\
                           
    \end{tabular}
    \end{center}    
\end{table}

\section{METHODS} \label{sec: method}

This section presents the final methodologies used for the Orca-Quest. Due to time constraints and evolving techniques, the data sampling for calculating the conversion gain, read noise, and dark current varied across different detectors. All data were collected in a controlled lab environment, maintaining consistent sampling techniques across each mode of the same detector. The data analysis methods remained consistent across all detectors. Given the individual pixel-wise amplification and readout in CMOS sensors, a pixel-wise analysis approach was adopted, characterizing each pixel independently. A custom Python wrapper was written for each detector to control various settings including readout mode, temperature, and exposure time. All images taken were saved as FITS files using the \textit{astropy} library\cite{astropy}.

\subsection{Conversion Gain} \label{sec: gain method}

The setup used to collect data for calculating the conversion gain involved collimating light from a 470 nm LED and passing it through an engineered diffuser to uniformly illuminate the entire detector. The detector captured 51 light frames at 40 exposure times linearly spaced from 0.001 seconds to 5 seconds for the standard and ultra-quiet modes, and from 0.001 seconds to 0.225 seconds for the photon number resolving mode. Additionally, 102 bias frames were collected before and after the light frames at the shortest allowed exposure time for each mode. While a true bias frame would be taken at a zero-second exposure to eliminate any signal generated by incident light, the exposure time was limited by the frame rate of the readout mode. The standard mode allowed for a minimum exposure time of 7.2 microseconds, while the slower ultra-quiet mode, and consequently the photon number resolving mode, had a minimum exposure time of 172.8 microseconds\cite{hama_tech_note}. This minimum exposure time was different for each detector.

A combined bias frame, created by stacking the 102 bias frames using the median value of each pixel across all frames, was subtracted from each light frame. For each exposure time, the variance and median value of each pixel in the light frames were used to create the variance and mean signal frames, respectively. Finally, the conversion gain for each pixel was calculated as the slope of a linear fit to a mean-variance plot of its signals and variances across all exposure times.

This entire process was repeated six times over three days to account for any environmental or power instabilities, and the conversion gain of each pixel was determined as the mean gain across the different datasets.

\subsection{Read noise} \label{sec: RN method}

The read noise was calculated as the standard deviation of each pixel across 1000 bias frames taken at the shortest allowed exposure time for each mode in a completely dark room with the lens cap attached to guarantee that no external light contributed to the signal. The standard deviation was then multiplied by the pixel's conversion gain to obtain the read noise in units of electrons. The data collection and analysis methods for read noise analysis were adapted from Ref. \citenum{Alarcon_2023}.

\subsection{Dark current} \label{sec: DC method}

The data to measure the dark current was also collected in a completely dark room with the lens cap attached to guarantee that no external light contributed to the signal. 21 dark frames were taken at a 60-second exposure time for each mode, along with 42 bias frames taken before and after the dark frames at the shortest allowed exposure time for each mode. As with the conversion gain, a median-stacked combined bias frame was created and subtracted from each dark frame. These bias-subtracted dark frames were then median-stacked to develop a combined dark frame. The signal of each pixel in this combined dark frame was divided by the exposure time and multiplied by its conversion gain to obtain the pixel's dark current in units of electrons per second. The data collection and analysis methods for dark current analysis were also adapted from Ref. \citenum{Alarcon_2023}.

The Orca-Quest is limited to operating at $-20^\circ$C with air cooling. For consistency, we tested dark current for all other detectors at the same air-cooled temperature of $-20^\circ$C, except for the Sona-11. Due to data collection prior to obtaining the Orca-Quest, the Sona-11 was tested at $-15^\circ$C and $-25^\circ$C.

\subsection{Quantum Efficiency} \label{sec: QE method}

The quantum efficiency analysis was conducted using the setup and scripts detailed in Ref. \citenum{QE}. The process is summarized as follows: light from a xenon lamp passes through a filter wheel into a monochromator, followed by an integrating sphere. Both the detector and a calibrated photodiode are placed equidistant from the integrating sphere. Measurements were taken simultaneously from the detector and photodiode for wavelengths ranging from 200 to 1100 nm in 10 nm increments, with 10 light images captured at each step. At each step, dark images were also taken and median-stacked to create a combined dark frame, which was then subtracted from each light image. The light images were median-stacked to create a combined light frame for each wavelength. The combined light frame was analyzed using the conversion gain and photodiode readings, following the procedure outlined in Ref. \citenum{QE}, to determine the quantum efficiency at each wavelength. The quantum efficiency calculations were performed as a global average instead of on a pixel-wise level.

\subsection{Linearity} \label{sec: lin method}

The linearity assessment utilizes the data from the conversion gain measurement described in Section \ref{sec: gain method}. It involves plotting the signal of each pixel from all the bias-subtracted combined signal frames against their corresponding exposure times and fitting a straight line to the data using linear regression. The fitting is only performed using signals below the saturation limit of the selected mode. The measure of each pixel's linearity is then determined by the average deviation of its signal from the line of best fit at each signal level. This value is expressed as a percentage, where a higher percentage indicates data that is more linear.

\section{RESULTS} \label{sec: results}

This section presents a detailed discussion of the characterization results of the Orca-Quest and its different modes. A summary of the characterization results for all tested detectors and modes is shown in Table \ref{tab: results}. Detailed results of the other detectors are given in Appendix \ref{sec: appendix}.

\begin{table}[h!]
    \caption{Summary of the global characterization results for all the tested modes for each sCMOS detector. More details about the performance of the Orca-Quest are provided in Sections \ref{sec: gain results}-\ref{sec: QE results} while those of the other detectors can be found in Appendix \ref{sec: appendix}. \underline{Note:} Due to limited availability and an early testing phase, the Sona-11 underwent a shorter data collection process compared to other detectors. Consequently, pixel-wise analysis could not be conducted and only aggregate results are presented here. Future analysis of the Sona-11 using the refined data sampling techniques may yield different results.}
    \label{tab: results}

    \begin{center}
    \begin{tabular}{c|c|cccc}
    
    Detector & Readout mode & \begin{tabular}[c]{@{}c@{}}Conversion gain\\ {[}e$^-$/ADU{]}\end{tabular} & \begin{tabular}[c]{@{}c@{}}Read noise\\ {[}e$^-${]}\end{tabular} & \begin{tabular}[c]{@{}c@{}}Dark current\\$@ -20^\circ$C\\ {[}e$^-$/s{]}  \end{tabular} & Linearity\\ 
    \hline
    \hline
    
    \multirow{3}{*}{\begin{tabular}[c]{@{}c@{}}Teledyne Prime\\95B (25 mm)\end{tabular}} 
                           & Balanced  
                           & $1.16 \pm 0.07$ & $2.4 \pm 0.5$ & $0.7 \pm 0.5$ & $>99.4\%$ \\
                           & Full Well  
                           & $2.3 \pm 0.2$ & $3.5 \pm 0.7$ & $2 \pm 1$ & $>99.4\%$ \\ 
                           & Sensitive 
                           & $0.60 \pm 0.03$ & $1.8 \pm 0.5$ & $0.4 \pm 0.3$ & $>99.0\%$ \\
                           \hline
    
    \begin{tabular}[c]{@{}c@{}}Andor Sona-11\\ (32 mm)\end{tabular}
                           & 16 bit  
                           & 1.30 & $3.0 \pm 0.6$ & \begin{tabular}[c]{@{}c@{}}$4.5 \pm 0.7$\\(@ $-25^\circ$C)\end{tabular} & $>99.5\%$ \\ 
                           \hline
    
    \begin{tabular}[c]{@{}c@{}}Andor Sona-6\\ Extreme\end{tabular}
                           & 12 bit  
                           & $0.4 \pm 0.1$ & $1.2 \pm 0.3$ & $0.2 \pm 0.4$ & $>99.4\%$ \\
                           \hline
                           
    \multirow{3}{*}{\begin{tabular}[c]{@{}c@{}}Ximea xiJ \\ MJ042MR-\\ GP-P6-BSI\end{tabular}}
                           & Standard  
                           & $1.7 \pm 0.3$  & $2.4 \pm 0.5$   & $0.5 \pm 0.4$   & $>98.6\%$ \\
                           & \begin{tabular}[c]{@{}c@{}}Correlated\\multiple\\sampling\end{tabular} & $3.1 \pm 0.4$  & $4.5 \pm 0.8$   & $2 \pm 1$   & $>98.8\%$ \\
                           \hline

    \multirow{3}{*}{\begin{tabular}[c]{@{}c@{}}Hamamatsu\\ Orca-Quest\\ C15550-20UP\end{tabular}}  
                           & Standard    
                           & $0.101 \pm 0.005$ & $0.37 \pm 0.09$ & $0.0067 \pm 0.0003$ & $>99.2\%$ \\
                           & Ultra-quiet 
                           & $0.108 \pm 0.004$ & $0.22 \pm 0.07$ & $0.0198 \pm 0.0008$ & $>99.2\%$ \\
                           & \begin{tabular}[c]{@{}c@{}}Photon number \\ resolving\end{tabular} 
                           & $1.04 \pm 0.04$   & --              & $0.0173 \pm 0.0007$ & $>97.7\%$ \\ 
                           
    \end{tabular}
    \end{center}    
\end{table}

Comparing these results to the manufacturer-reported values in Table \ref{tab: detectors}, it is evident that, except for the Sona-11, the calculated read noise, dark current, and linearity values for all detectors align with the manufacturer's specifications within uncertainties. Notably, the dark current of the Orca-Quest and the linearity of the Ximea xiJ exceed expectations. In contrast, the Sona-11 underperforms in read noise and dark current but shows consistent linearity. This is likely because the Sona-11 data was collected early in the project, before more detailed data sampling methodologies were developed. Consequently, pixel-wise testing could not be conducted. Further testing of the Sona-11 with the now-established methodologies is necessary to verify its results.

Among the tested detectors, the Orca-Quest significantly outperforms in read noise and dark current measurements by an order of magnitude, while the Prime 95B, Sona-11, and Sona-6 show the highest linearity.

\subsection{Conversion Gain} \label{sec: gain results}

\begin{figure}[h!]
  \centering
  \includegraphics[width=0.9\columnwidth]{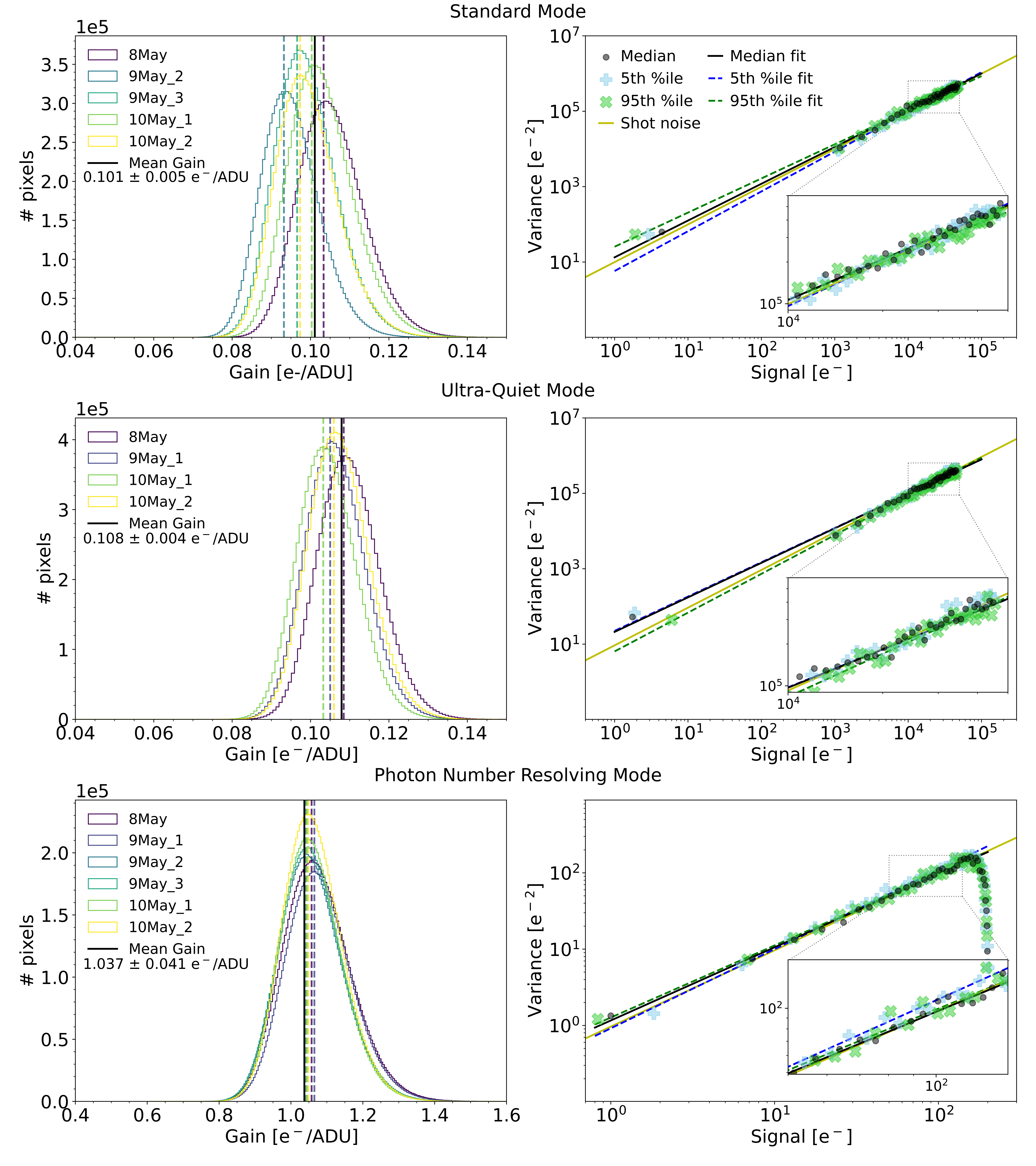}
\caption{Results of the Orca-Quest's conversion gain calculations for three modes: standard (top), ultra-quiet (middle), and photon number resolving (bottom). \textit{Left:} Histograms showing the distribution of conversion gain values across all pixels in the Orca-Quest. Dashed lines indicate the median gain for each dataset, while the solid black line represents the mean gain across all datasets. \textit{Right:} Mean-variance log plots for pixels with median gain, 5th percentile gain, and 95th percentile gain. The solid yellow line represents the median shot noise. The inset plot zooms in on the high signal regime. The photon number resolving mode saturates at about 200 e$^-$, causing the drop in variance in the bottom right figure.}
  \label{fig: gain}
\end{figure}

The left plots in Figure \ref{fig: gain} display the per-pixel gain distribution for each mode of the Orca-Quest and the differences in gain distribution across various datasets. The right plots in the same figure present mean-variance plots for pixels with median gain, 5th percentile gain, and 95th percentile gain.

Each mode displays normally distributed conversion gain values. The standard and ultra-quiet modes have narrower gain distributions compared to the photon counting mode. However, the standard mode exhibits greater instability in the shape and location of the distribution across different datasets, with a $3.1\%$ deviation in the median gain, compared to $1.8\%$ in the ultra-quiet mode and $0.86\%$ in the photon number resolving mode. It is unclear whether this observed instability between different datasets is due to the detector itself or external factors such as fluctuations in the light source's power supply. More data collected over a longer period would help identify the source of this variation. Due to time constraints, we were unable to gather the necessary data to investigate this variation for any of the other detectors.

The mean-variance plots for each mode indicate that the performance of each pixel is roughly consistent with the Poisson-distributed shot noise, calculated as $\sigma^2 = g/\Bar{x}$, where $\sigma^2$ is variance, $g$ is median gain, and $\Bar{x}$ is the mean signal. Since both the 5th and 95th percentile pixels also align with the median shot noise line, it can be concluded that the detector's performance is limited by shot noise and is not significantly affected by read noise or FPN. Negligible FPN contributions were also observed in the Prime 95B, Sona-6, and Ximea xiJ detectors. The FPN contribution of the Sona-11 could not be identified due to a lack of sufficient data. Interestingly, the FPN contributions of the Sony IMX455 and IMX411 sensors in the QHY600M Pro and QHY411M detectors, shown in Figure 8 of Ref. \citenum{Alarcon_2023}, were non-negligible at $0.55 \pm 0.07\%$ and $0.31 \pm 0.02\%$ respectively. This suggests that higher-end commercial sCMOS detectors exhibit much less pixel-to-pixel variation compared to lower-cost models.

\subsection{Read noise} \label{sec: RN results}

\begin{figure}[h]
  \centering
  \includegraphics[width=0.65\columnwidth]{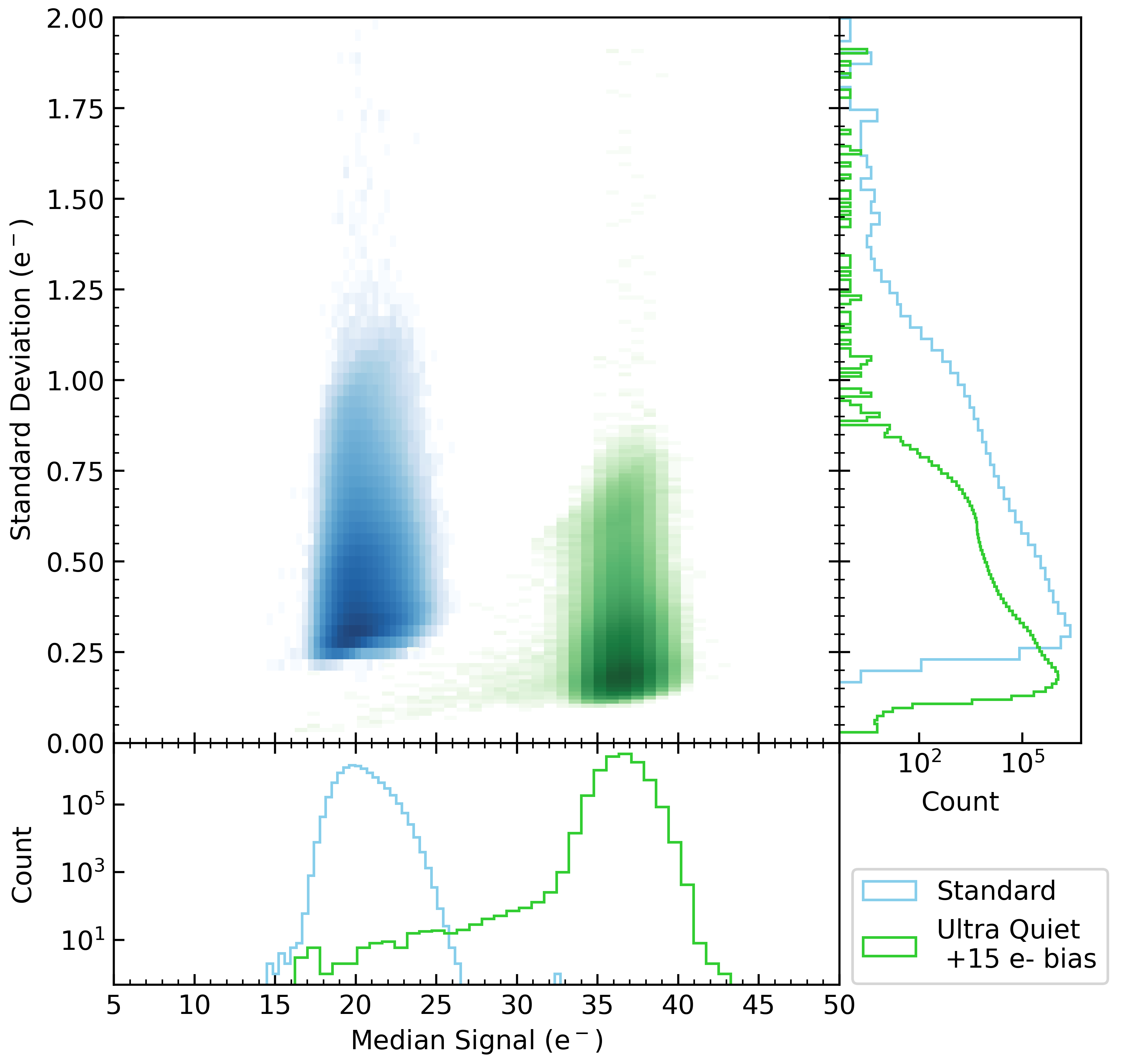}
  \caption{A 2-dimensional histogram showing the distribution of each pixel's read noise (standard deviation) and bias (median signal) levels in the Orca-Quest for the standard (blue) and ultra-quiet (green) modes. The bias of the ultra-quiet mode has been artificially shifted 15 e$^-$ to the right to prevent overlap and improve data visualization. The side 1-dimensional histograms display the read noise distribution on the right and the bias distribution on the bottom.}
  \label{fig: RN}
\end{figure}

The per-pixel bias and read noise values of the Orca-Quest's standard and ultra-quiet modes are shown in Figure \ref{fig: RN}. Note that the ultra-quiet bias has been artificially shifted by 15 e$^-$ to improve data visualization.

As expected, the ultra-quiet mode, which is designed to minimize noise, exhibits lower read noise values and a narrower distribution compared to the standard mode. Notably, the bias distribution in the standard mode is approximately normal, whereas the ultra-quiet mode is skewed towards a lower bias. This skew only occurs after multiplication with the gain map and does not affect the sensor performance since the bias can be subtracted away. Thus, this skew is insignificant in real-world applications. Despite these differences in distribution shapes, both modes demonstrate a stable and well-defined average signal ranging between 17 and 25 e$^-$. The read noise is low for both modes, averaging at $0.37 \pm 0.09$ e$^-$ for the standard mode and $0.22 \pm 0.07$ e$^-$ for the ultra-quiet mode. The photon number resolving mode was excluded from read noise testing because it is a calibrated version of the ultra-quiet mode, applying real-time signal correction to calculate the integer number of incident photons\cite{hama_tech_note}. As a result, read noise is not defined for this mode.

Interestingly, none of the detectors tested in this study exhibit the dispersion in the standard deviation vs. mean signal plot observed for the Sony IMX411 sensor, used in the QHY411M detector, as shown in Figure 3 of Ref. \citenum{Alarcon_2023}. This suggests that higher-end commercial sCMOS detectors have fewer faulty pixels compared to their lower-cost counterparts. A detailed characterization of salt-and-pepper pixels would help strengthen this finding.

Our results show that the read noise levels in all the tested commercial sCMOS detectors are significantly lower than those in comparable CCDs.

\subsection{Dark current} \label{sec: DC results}

\begin{figure}[h]
  \centering
  \includegraphics[width=0.9\columnwidth]{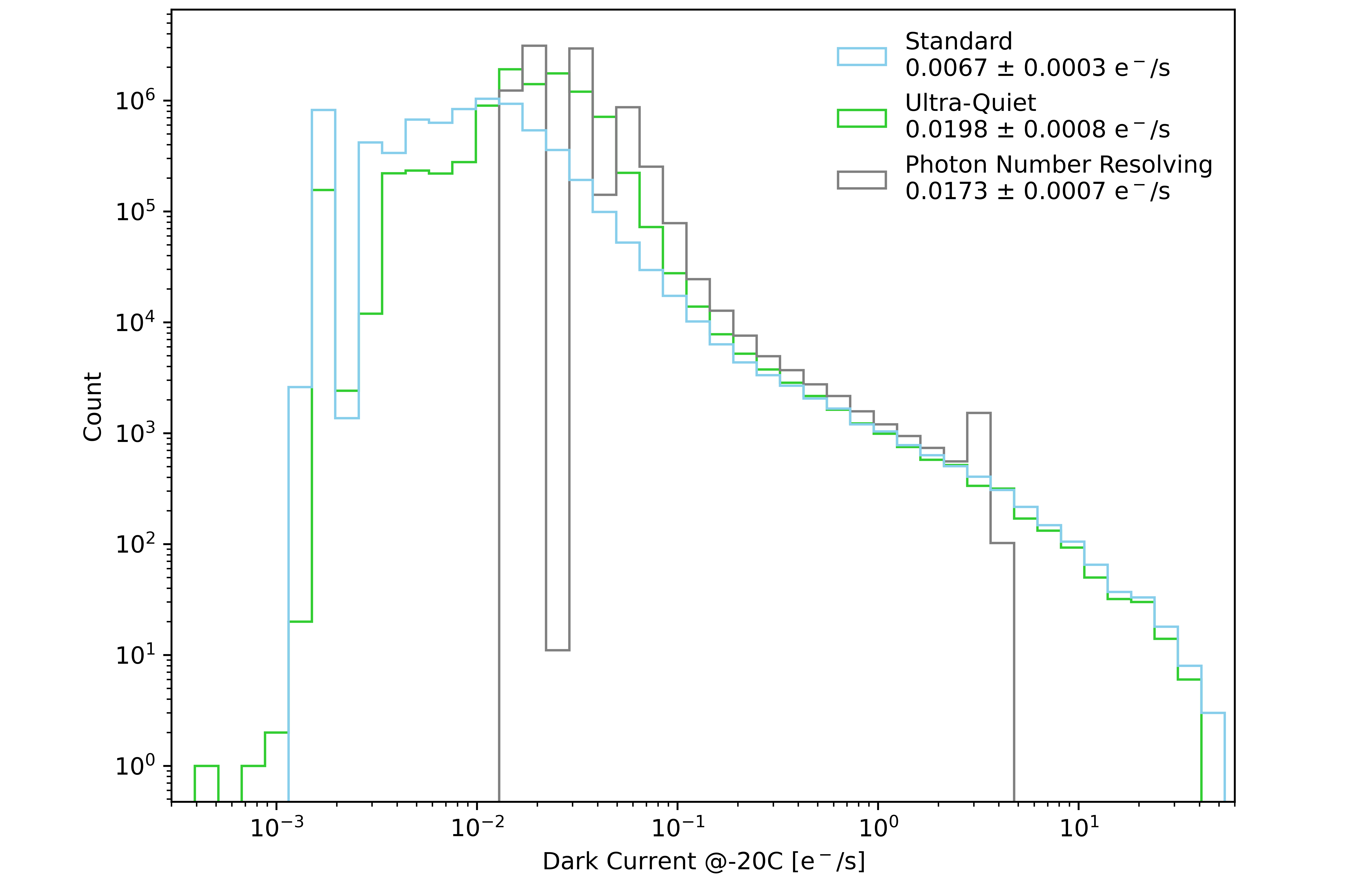}
  \caption{Histograms showing the distribution of each pixel's dark current level in the Orca-Quest for the standard (blue), ultra-quiet (green), and photon number resolving (grey) modes at $-20^\circ$C.}
  \label{fig: DC}
\end{figure}

The per-pixel dark current performance of the Orca-Quest across all three modes can be seen in Figure \ref{fig: DC}. Since dark current (DC) is determined by the thermal fluctuations in the silicon and is not affected by the readout mechanism, the histograms for all the standard and ultra-quiet modes are very similar.

The slight bump and sharp cut-off in the dark current of the photon number resolving mode near 3 e$^-$/s are due to the detector reaching its 200 e$^-$ saturation limit. The abrupt drop near 0.02 e$^-$/s and the dip around 0.03 e$^-$/s correspond to signals of 1 e$^-$ and 2 e$^-$, respectively, divided by the 60-second exposure time. This is because the photon number resolving mode only outputs integer values.

The Orca-Quest exhibited the lowest dark current levels among all tested detectors, averaging $0.0067 \pm 0.0003$ e$^-$/s in standard mode and $0.0198 \pm 0.0008$ e$^-$/s in ultra-quiet mode. Except for the Ximea xiJ, all tested detectors offer liquid cooling, which can reduce dark current levels. However, our study focused on air-cooled dark current levels, revealing that most commercial sCMOS detectors still showed higher dark current levels than equivalent CCDs at $-20^\circ$C. Notably, the high-sensitivity Orca-Quest was the only detector achieving lower dark current levels than CCDs with air cooling alone.

\subsection{Quantum Efficiency} \label{sec: QE results}

\begin{figure}[h!]
  \centering
  \includegraphics[width = 0.75\columnwidth]{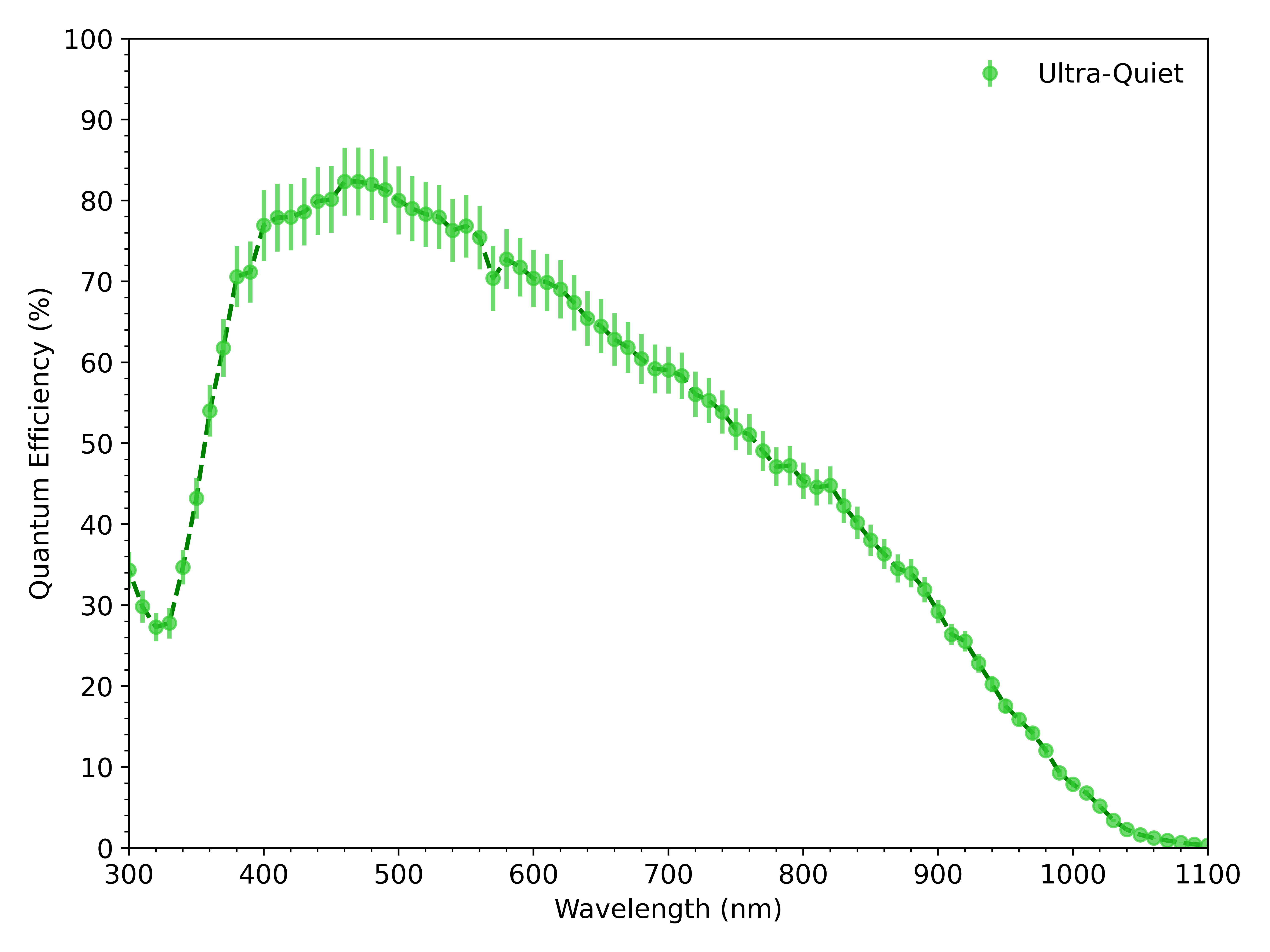}
  \caption{Global combined quantum efficiency for the Orca-Quest's sensor and front window in ultra-quiet mode. \underline{Note:} This figure has been revised to address an issue identified in the test setup. It now includes only the data deemed reliable.}
  \label{fig: QE}
\end{figure}

Figure \ref{fig: QE} shows the combined quantum efficiency curve of the Orca-Quest's sensor and front window. Unlike other tests, the quantum efficiency was characterized globally rather than pixel-wise. The peak quantum efficiency of $82\% \pm 4\%$ occurs at 470 nm. To determine the quantum efficiency of the sensor alone, it is necessary to normalize this curve using the transmission curve of the detector's front window. Since quantum efficiency is an inherent characteristic of the sensor and should remain consistent regardless of the readout mode, data was collected only for the ultra-quiet mode. Notably, this plot aligns closely with the sensor-only quantum efficiency curve illustrated in Figure 2-3 of Ref. \citenum{hama_tech_note} within uncertainties. The vendor curve indicates a drop to $0\%$ efficiency for wavelengths below 300 nm. We have yet to measure the quantum efficiency in this regime accurately. Due to time constraints, we were only able to measure the quantum efficiency curve of the Orca-Quest.

\subsection{Linearity} \label{sec: lin results}

\begin{figure}[h!]
  \centering
  \includegraphics[width = 0.75\columnwidth]{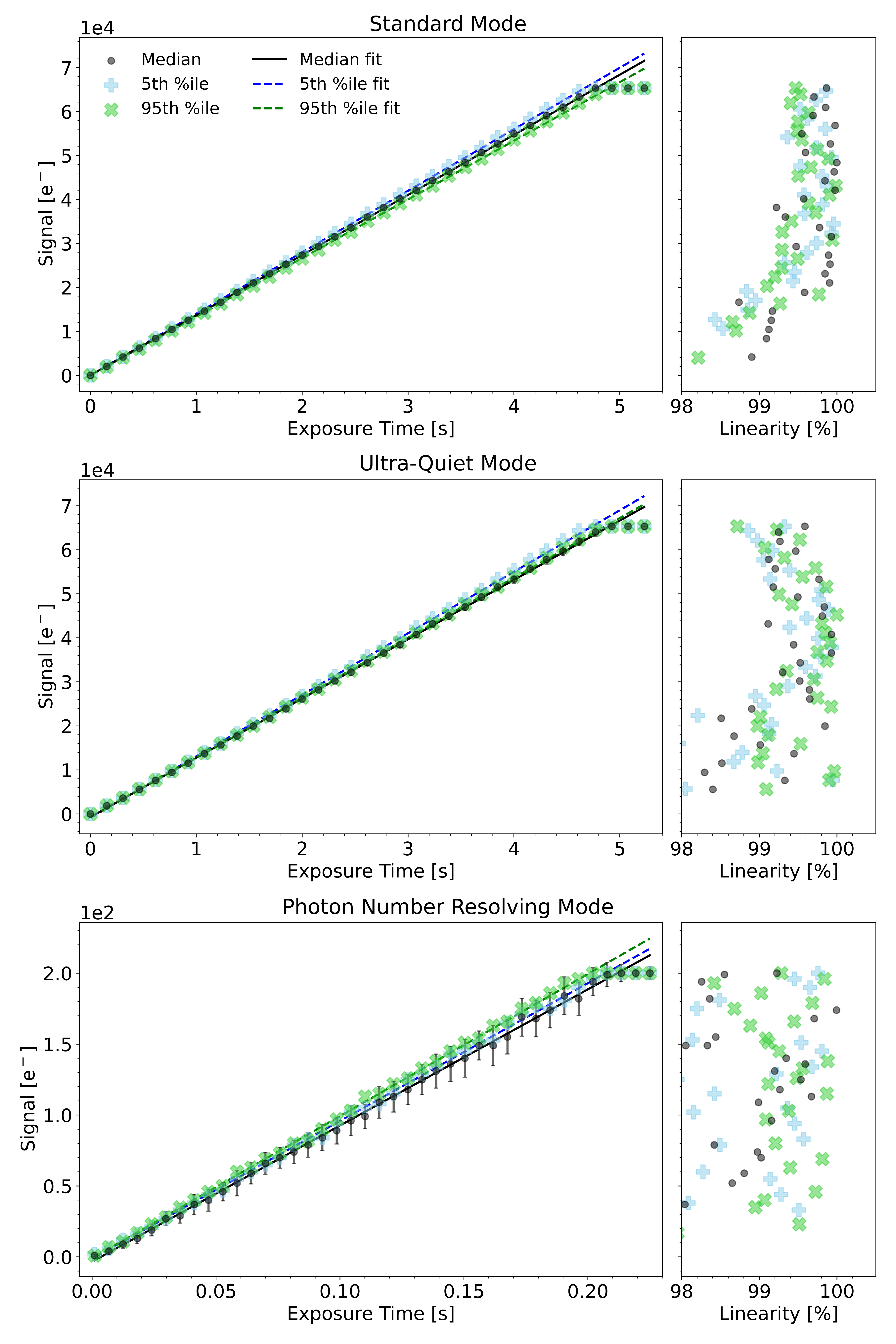}
  \caption{Linearity plots for the Orca-Quest's three modes: standard (top), ultra-quiet (middle), and photon number resolving (bottom). Each plot shows the performance of the pixels with the median gain (black), 5th percentile gain (blue), and 95th percentile gain (green). The residual plot on the right illustrates how close each point is to perfect linearity, expressed as a percentage. The signal plateaus around 65,000 e$^-$ in the standard and ultra-quiet modes, and around 200 e$^-$ in the photon number resolving mode, due to the detector saturating.}
  \label{fig: lin}
\end{figure}

Figure \ref{fig: lin} shows the linearity of the pixels with median gain, 5th percentile gain, and 95th percentile gain across the three modes tested for the Orca-Quest. The standard and ultra-quiet modes exhibit excellent linearity, each exceeding $99\%$ with minimal variance among the pixels. However, the photon number resolving mode, displaying linearity just below $98\%$, has more significant variations in linearity between the selected pixels. As this mode is a calibration of the more consistent ultra-quiet mode, this disparity between pixels and loss of linearity is most likely due to issues in the calibration algorithm.

\newpage

\newpage

\section{CONCLUSION} \label{sec: conc}

In this paper, we outlined a range of high-end commercial sCMOS detectors from various manufacturers, presenting our in-lab characterizations of their conversion gain, read noise, dark current, and linearity. These detectors were tested during short demo periods borrowed from their respective manufacturers. Notably, the Orca-Quest was purchased after its initial demo period to investigate its photon number resolving capabilities, so we focused primarily on its detailed characterization results, providing only summarized results for the other detectors.

Unlike CCDs, which use a single global amplifier with a shift register, sCMOS pixels have individual readout electronics, requiring each pixel to be tested as an independent detector. Historically, this led to high fixed pattern noise in CMOS detectors, but we found negligible fixed pattern noise in almost all the detectors we analyzed pixel-wise. Performance was shot noise limited in the middle and high signal regimes and read noise limited in the low signal regime. Due to time constraints and evolving methodology, pixel-wise analysis could not be performed for the Sona-11. We observed variations in the conversion gain values of the Orca-Quest across datasets taken over a few hours to a few days. However, it is unclear whether this is inherent to the detector or caused by external factors. Performance across multiple datasets could not be tested in any other detector.

CMOS technology inherently offers improved read noise over CCD technology, and we measured exceptionally low noise levels across all detectors. The Orca-Quest, in particular, exhibited sub-electron noise levels. Although multiplication with pixel-wise gain values resulted in a larger spread in the bias signal levels in some detectors, this effect is insignificant as these signals are subtracted in real-world applications.

The dark current level was under 1 e$^-$/s at an air-cooled temperature of $-20^\circ$C in most detectors, demonstrating excellent performance, although still short of CCDs. The Orca-Quest outperformed other tested detectors and even surpassed CCDs in dark current performance. It is also important to note that the standard deviation in the global dark current level, after multiplication with the pixel-wise gain map, was high, indicating a large variation across different pixels. The implications of this have yet to be investigated.

Quantum efficiency measurement could only be performed for the Orca-Quest, achieving a combined (sensor plus front window) peak efficiency of $82\%$ at 470 nm, gradually decreasing to $0\%$ at 1100 nm and extending to at least 300 nm in the UV regime. The reported peak quantum efficiency of all other detectors is at least $91\%$, generally reaching $95\%$, comparable to CCD performance.

The measured linearity of all detectors was above $97.7\%$, with most detectors showing less than a $1\%$ deviation from linearity, matching CCD performance.

Our results show that the modern high-end sCMOS detector technology is on par with, and in some cases surpasses, CCD technology, making sCMOS a highly competitive alternative for scientific optical astronomy. Due to their lower read noise, faster frame rates, and greater dynamic range, sCMOS detectors are better suited than CCDs for applications such as wavefront sensing and high-energy UV and soft X-ray detection. sCMOS technology is already being employed in instruments such as the ArgusSpec spectroscopic system for transient detection\cite{argusspec} and is planned for future instruments like the Cosmological Advanced Survey Telescope for Optical and Ultraviolet Research (CASTOR)\cite{CASTOR}.

Future directions for this project include on-sky photometric testing, further quantum efficiency characterization, evaluation of additional sCMOS detectors, charge persistence testing, and further investigation of dataset variations in the Orca-Quest and other detectors. We also aim to retake data with the borrowed detectors, particularly the Sona-11, if possible.

\acknowledgments \label{sec: acknowledgements}
We thank the seed funding from the Dunlap Institute for Astronomy and Astrophysics at the University of Toronto that helped enable this work. The Dunlap Institute is funded through an endowment established by the David Dunlap family and the University of Toronto. We also thank Teledyne Photometrics and Oxford Instruments for providing the Prime 95B, Sona-11, and Sona-6 detector demos for testing and characterization purposes.

\bibliography{report} \label{sec: bibliography}
\bibliographystyle{spiebib}

\appendix
\section{OTHER DETECTORS} \label{sec: appendix}
This appendix provides the plots showing the pixel-wise performances of the Prime 95B, Sona-11, Sona-6, and Ximea xiJ detectors. Since these detectors were borrowed for short demo periods, they could not be tested at the same level of detail as the Orca-Quest.

\subsection{Teledyne Prime 95B (25 mm)} \label{sec: teledyne}

\begin{figure}[h!]
  \centering
  \includegraphics[width=0.9\columnwidth]{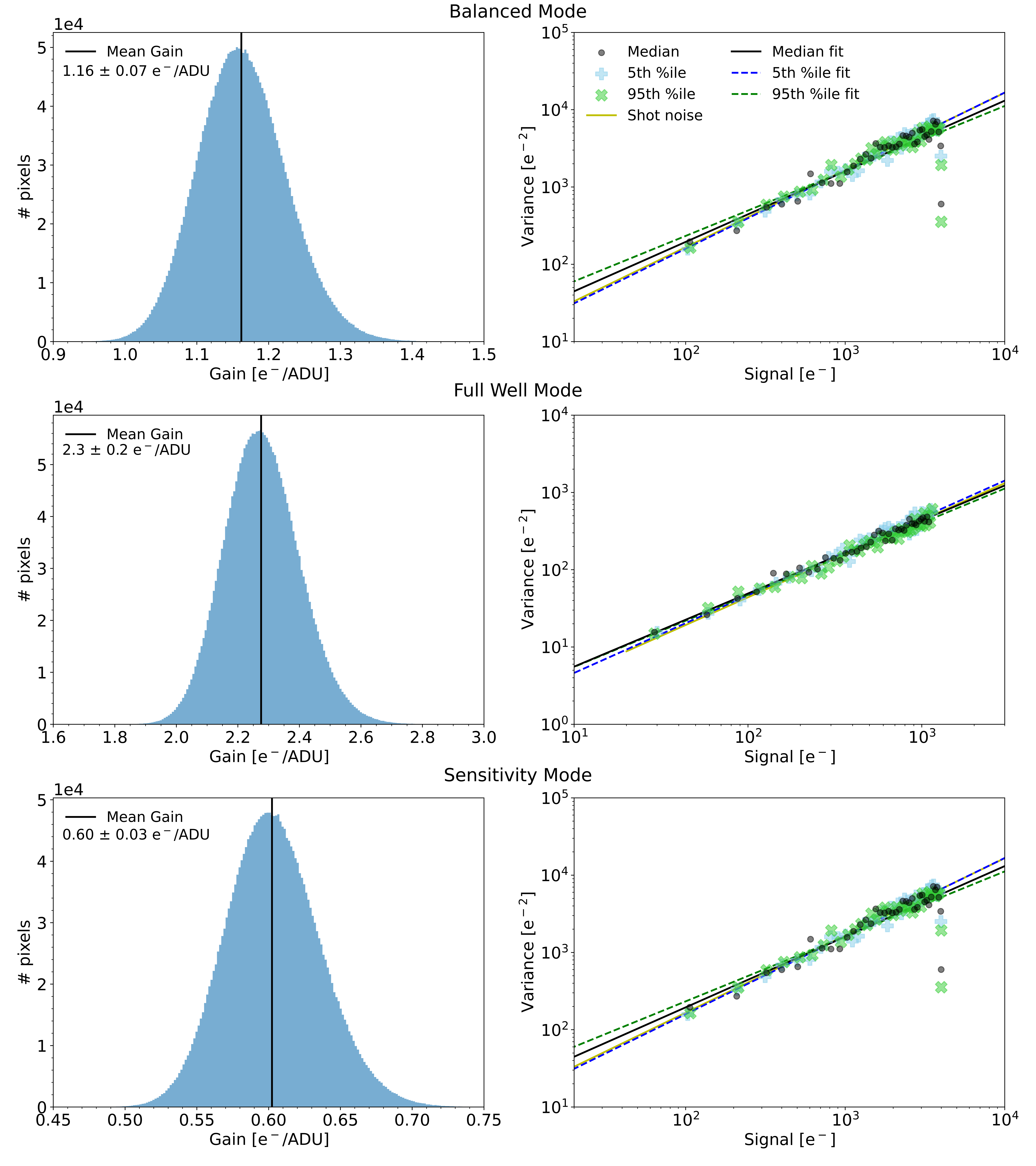}
\caption{Results of the Prime 95B's conversion gain calculations for three modes: balanced (top), full well (middle), and sensitivity (bottom). \textit{Left:} Histograms showing the distribution of conversion gain values across all pixels in the Prime 95B. The solid black line represents the mean gain across all pixels. \textit{Right:} Mean-variance log plots for pixels with median gain, 5th percentile gain, and 95th percentile gain. The solid yellow line represents the median shot noise.}
  \label{fig: gain_prime}
\end{figure}

\begin{figure}[h!]
  \centering
  \includegraphics[width=0.65\columnwidth]{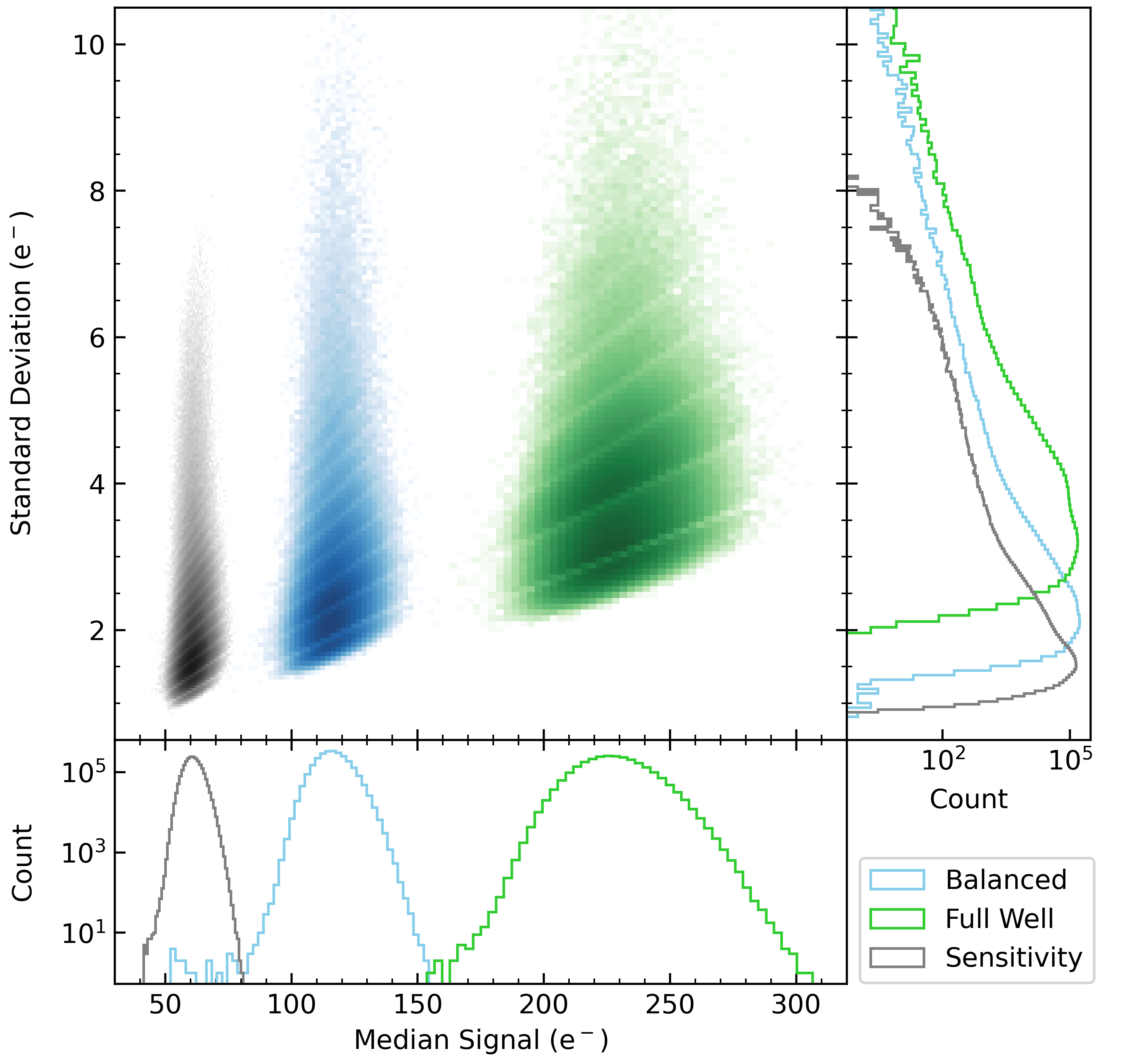}
  \caption{A 2-dimensional histogram showing the distribution of each pixel's read noise (standard deviation) and bias (median signal) levels in the Prime 95B for the balanced (blue), full well (green), and sensitivity (grey) modes. The side 1-dimensional histograms display the read noise distribution on the right and the bias distribution on the bottom.}
  \label{fig: RN_prime}
\end{figure}

\begin{figure}[h!]
  \centering
  \includegraphics[width=0.8\columnwidth]{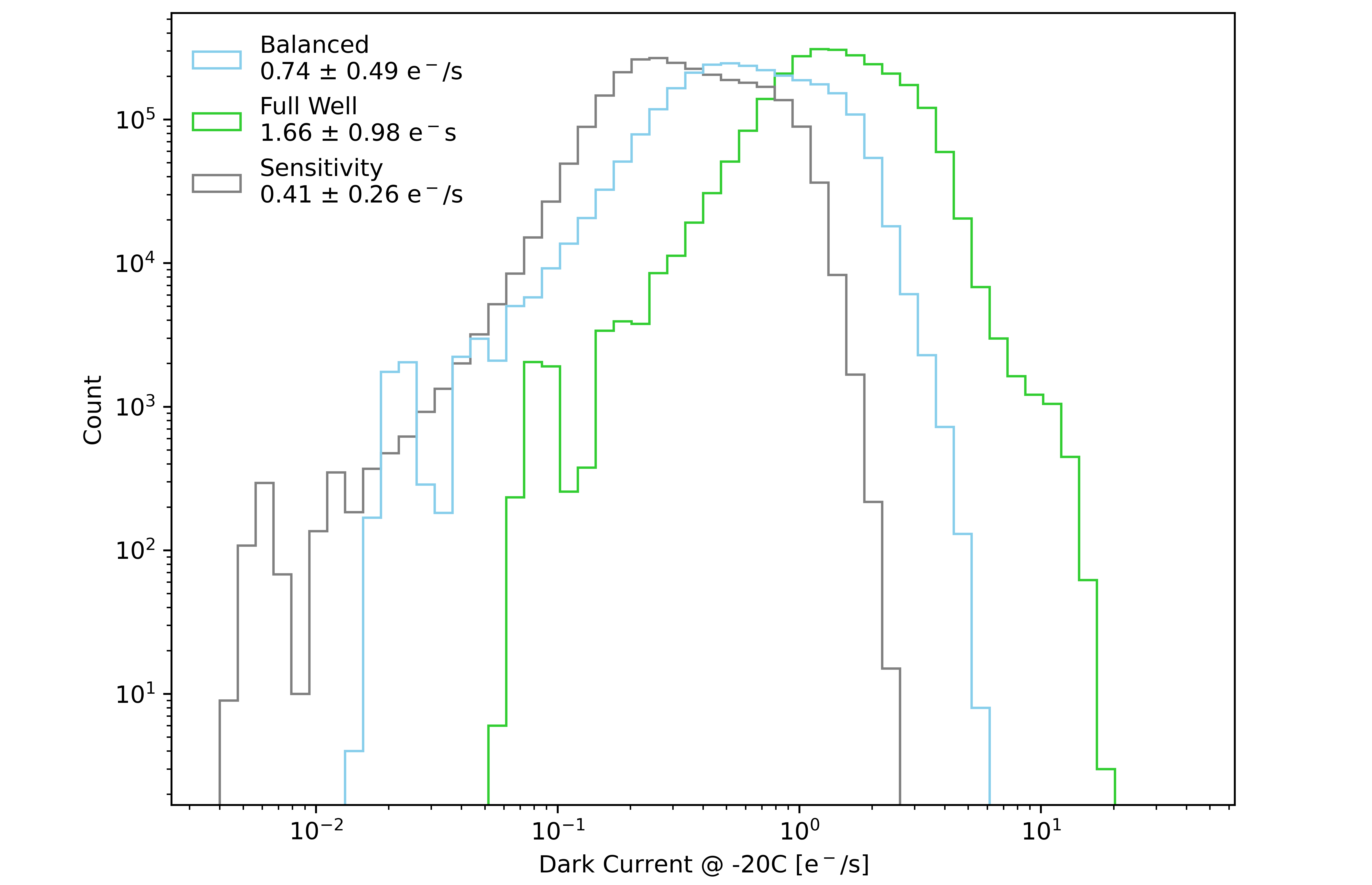}
  \caption{Histograms showing the distribution of each pixel's dark current level in the Prime 95B for the balanced (blue), full well (green), and sensitivity (grey) modes at $-20^\circ$C.}
  \label{fig: DC_prime}
\end{figure}

\begin{figure}[h!]
  \centering
  \includegraphics[width = 0.78\columnwidth]{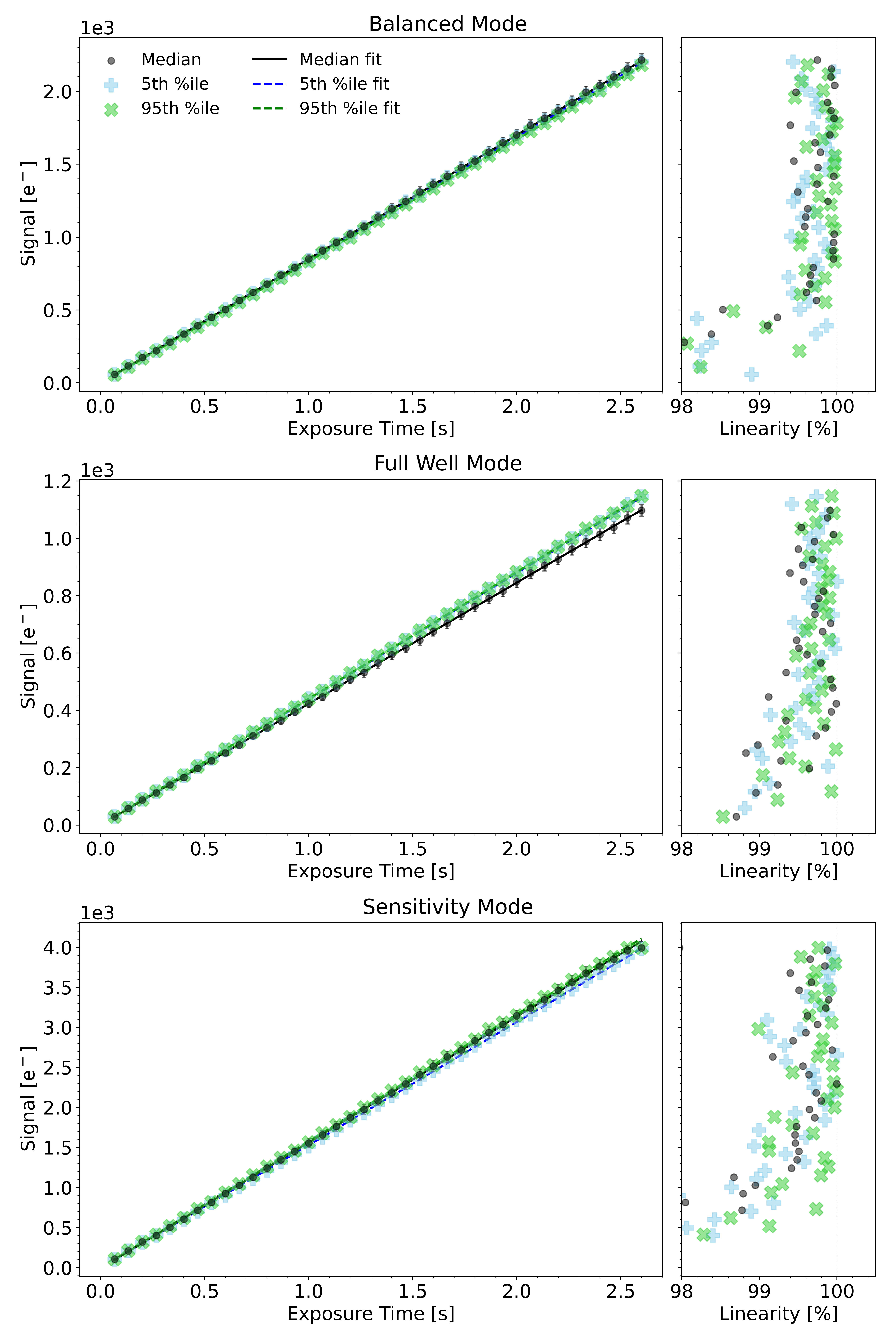}
  \caption{Linearity plots for the Prime 95B's three modes: balanced (top), full well (middle), and sensitivity (bottom). Each plot shows the performance of the pixels with the median gain (black), 5th percentile gain (blue), and 95th percentile gain (green). The residual plot on the right illustrates how close each point is to perfect linearity, expressed as a percentage.}
  \label{fig: lin_prime}
\end{figure}

\newpage

\subsection{Andor Sona-11 (32 mm)} \label{sec: andor 11}

\begin{figure}[h]
  \centering
  \includegraphics[width=0.8\columnwidth]{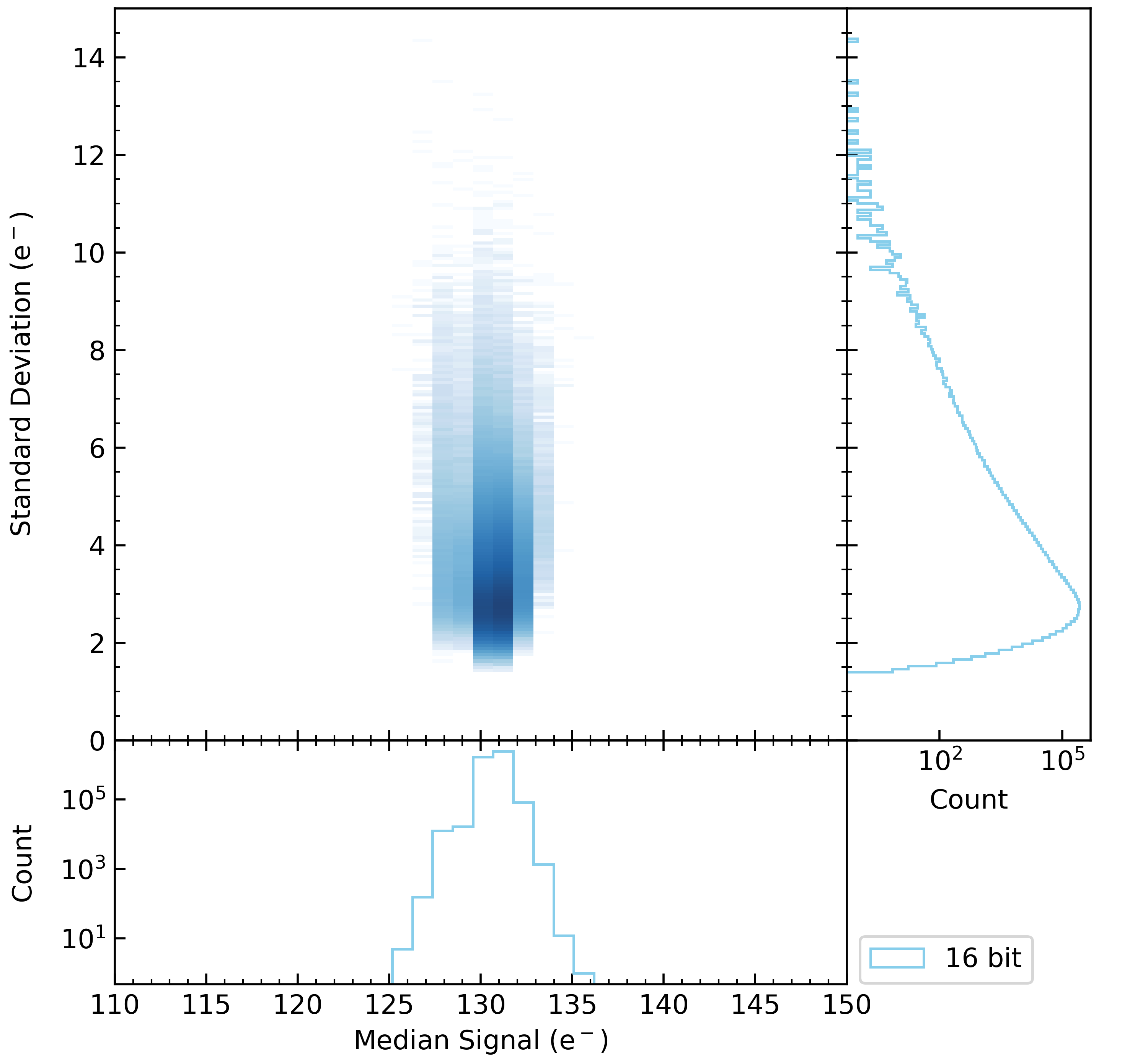}
  \caption{A 2-dimensional histogram showing the distribution of each pixel's read noise (standard deviation) and bias (median signal) levels in the Sona-11's 16-bit mode. The side 1-dimensional histograms display the read noise distribution on the right and the bias distribution on the bottom. \underline{Note:} The pixel-wise conversion gain for the Sona-11 could not be calculated. Therefore, all pixels have been multiplied by the global average gain value in this plot. Consequently, this representation does not reflect the true pixel-wise read noise and bias distribution.}
  \label{fig: RN_11}
\end{figure}

\begin{figure}[h!]
  \centering
  \includegraphics[width=\columnwidth]{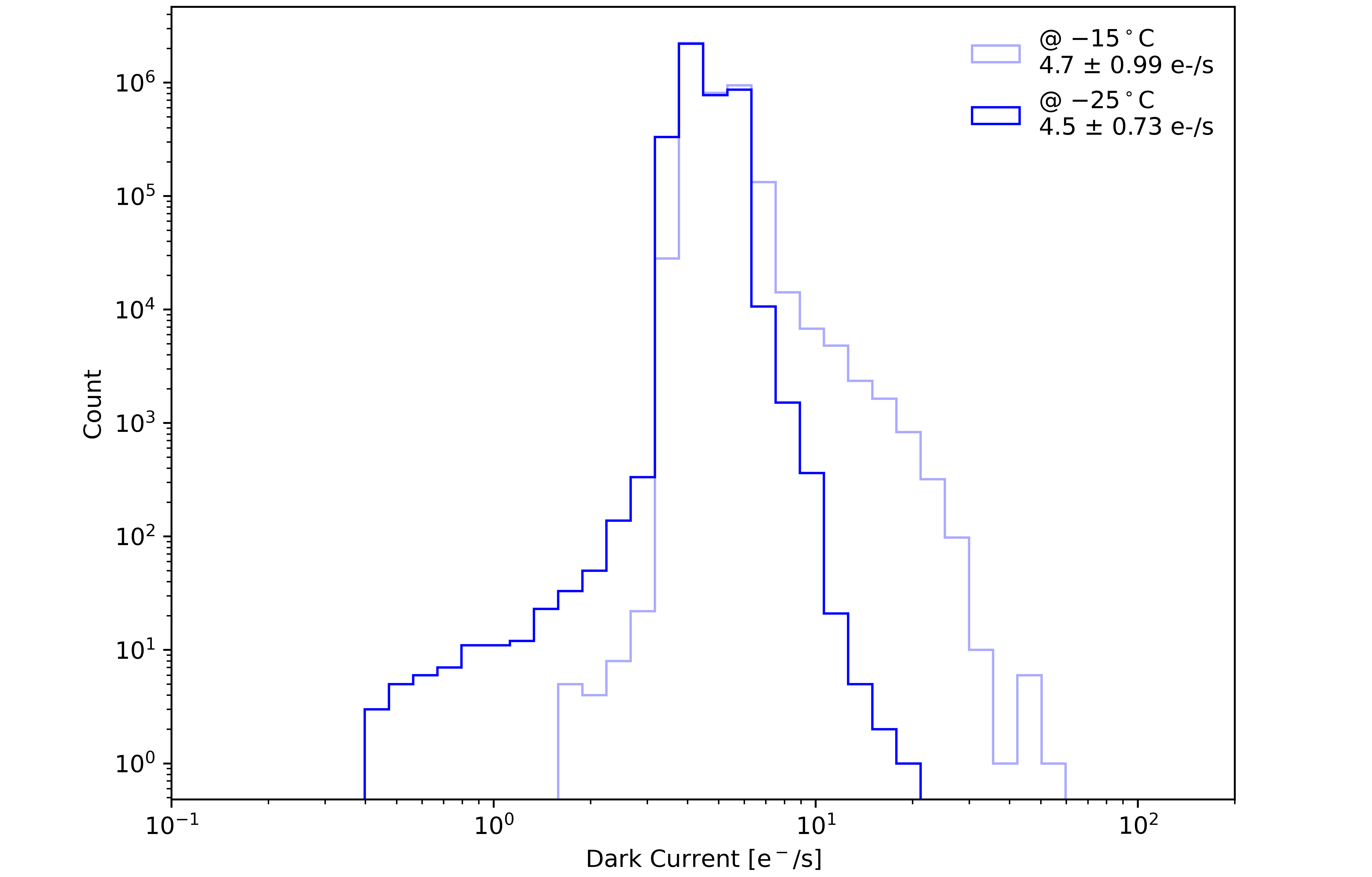}
  \caption{Histograms showing the distribution of each pixel's dark current level in the Sona-11 for the 16-bit mode at $-15^\circ$C (light blue) and $-25^\circ$C (dark blue). \underline{Note:} The pixel-wise conversion gain for the Sona-11 could not be calculated. Therefore, all pixels have been multiplied by the global average gain value in this plot. Consequently, this representation does not reflect the true pixel-wise dark current distributions.}
  \label{fig: DC_11}
\end{figure}

\begin{figure}[h!]
  \centering
  \includegraphics[width = 0.85\columnwidth]{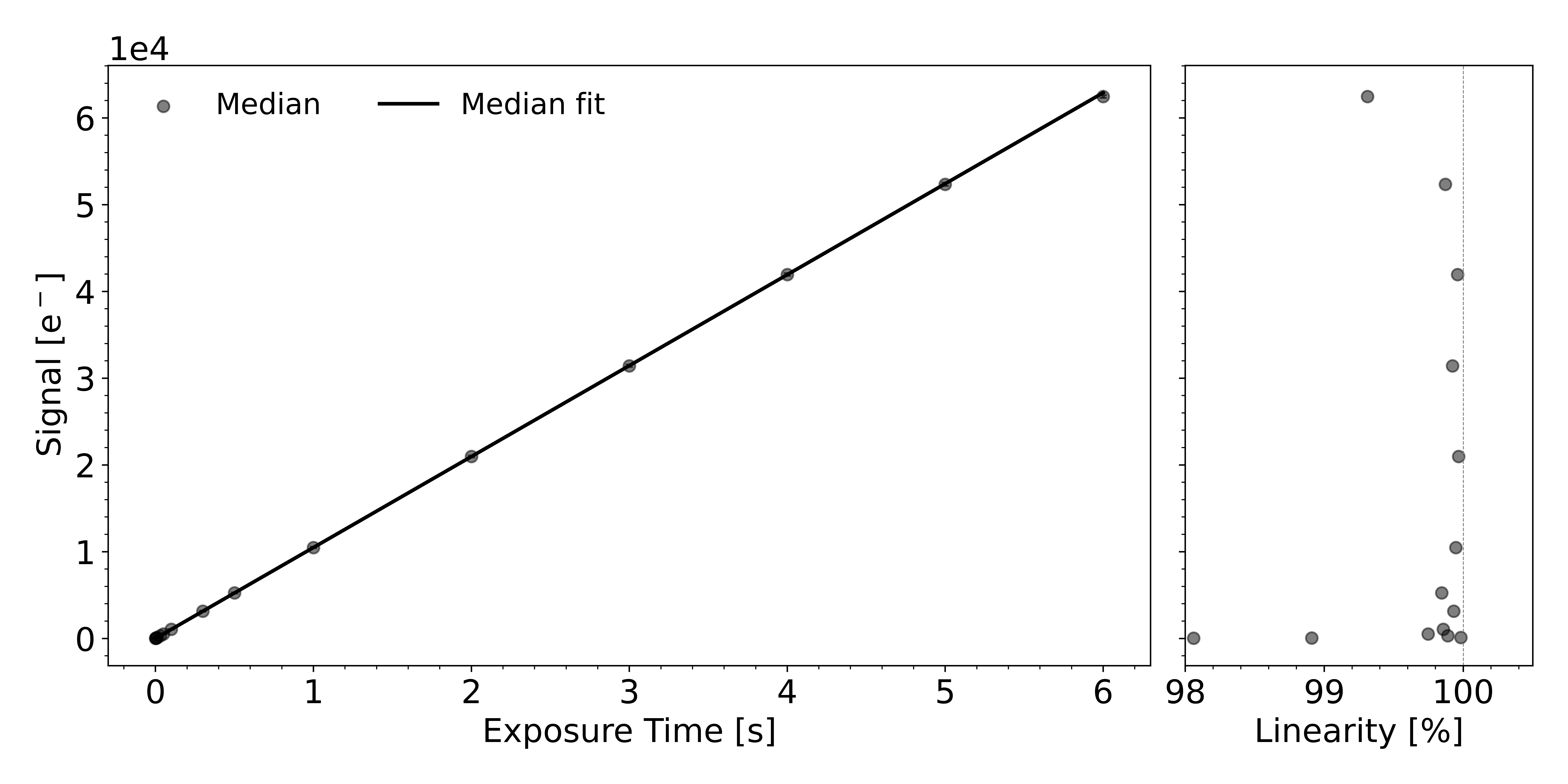}
  \caption{Linearity plot for the Sona-11's 16-bit mode. The plot shows the performance of a single pixel multiplied by the global gain. The residual plot on the right illustrates how close each point is to perfect linearity, expressed as a percentage.}
  \label{fig: lin_11}
\end{figure}

\newpage

\subsection{Andor Sona-6 Extreme}  \label{sec: andor 6}

\begin{figure}[h!]
  \centering
  \includegraphics[width=0.85\columnwidth]{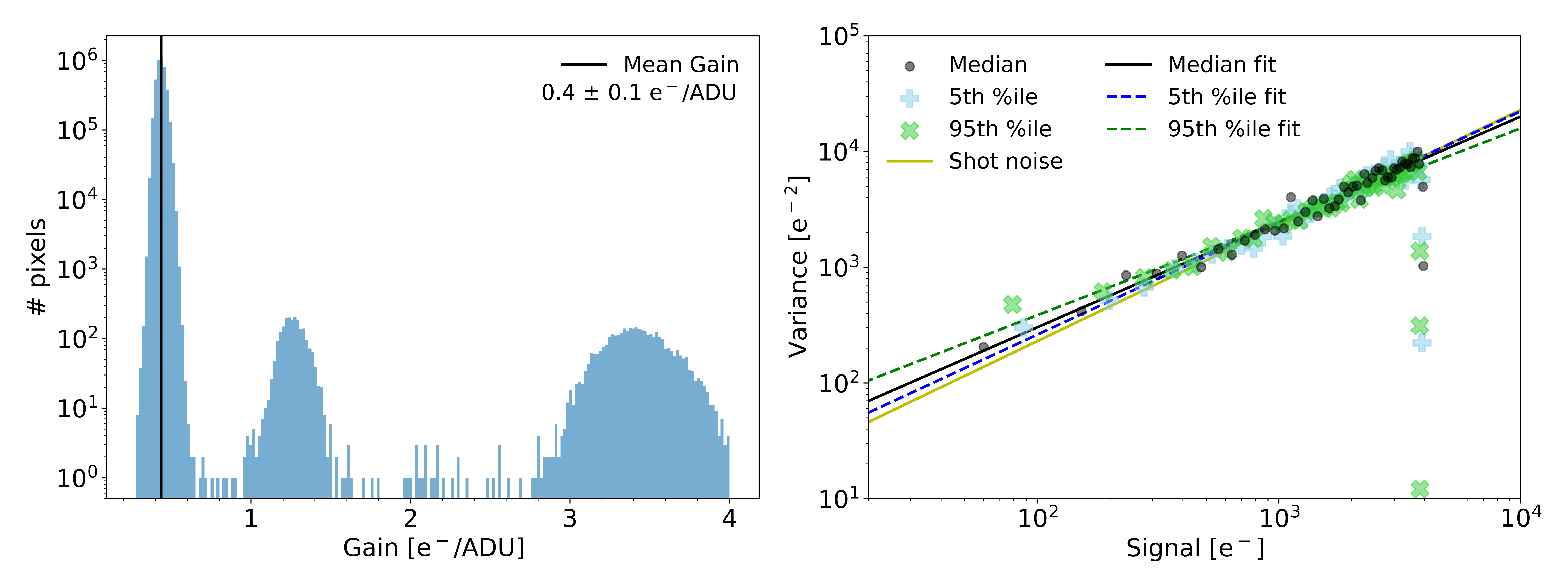}
\caption{Results of the Sona-6's conversion gain calculations for the 12-bit mode. \textit{Left:} Histogram showing the distribution of conversion gain values across all pixels in the Sona-6. The solid black line represents the mean gain across all pixels. \textit{Right:} Mean-variance log plots for pixels with median gain, 5th percentile gain, and 95th percentile gain. The solid yellow line represents the median shot noise. \underline{Note:} The gain distribution of the Sona-6 shows three distinct gain groupings of pixels. More data is needed to test if this is just a result of bad data or cased due to the detector itself.}
  \label{fig: gain_6}
\end{figure}

\begin{figure}[h]
  \centering
  \includegraphics[width=0.6\columnwidth]{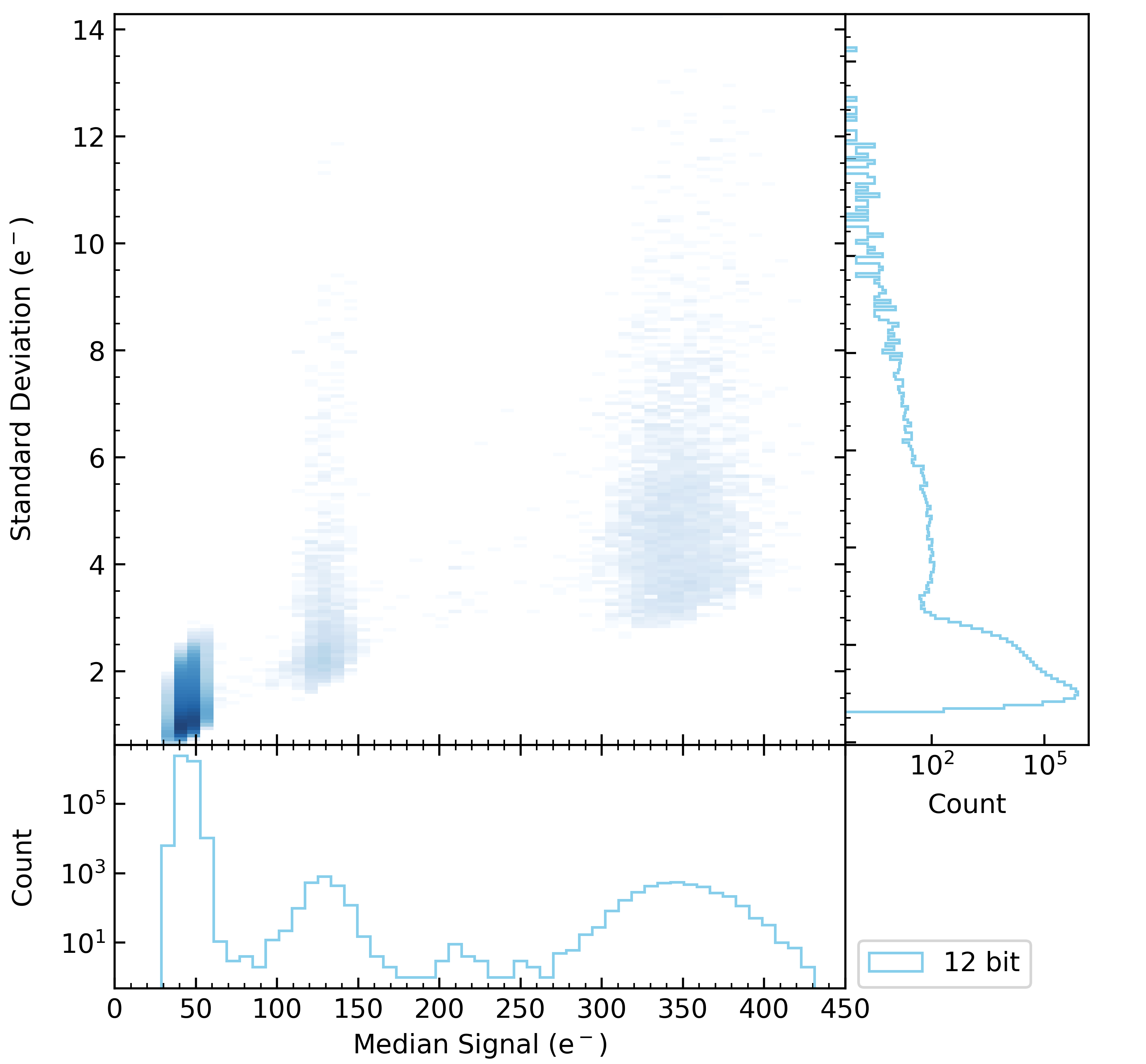}
  \caption{A 2-dimensional histogram showing the distribution of each pixel's read noise (standard deviation) and bias (median signal) levels in the Sona-6's 12-bit mode. The side 1-dimensional histograms display the read noise distribution on the right and the bias distribution on the bottom. \underline{Note:} The multiple bias grouping are caused due to the peaks in the gain map which are almost identical in shape.}
  \label{fig: RN_6}
\end{figure}

\begin{figure}[h]
  \centering
  \includegraphics[width=\columnwidth]{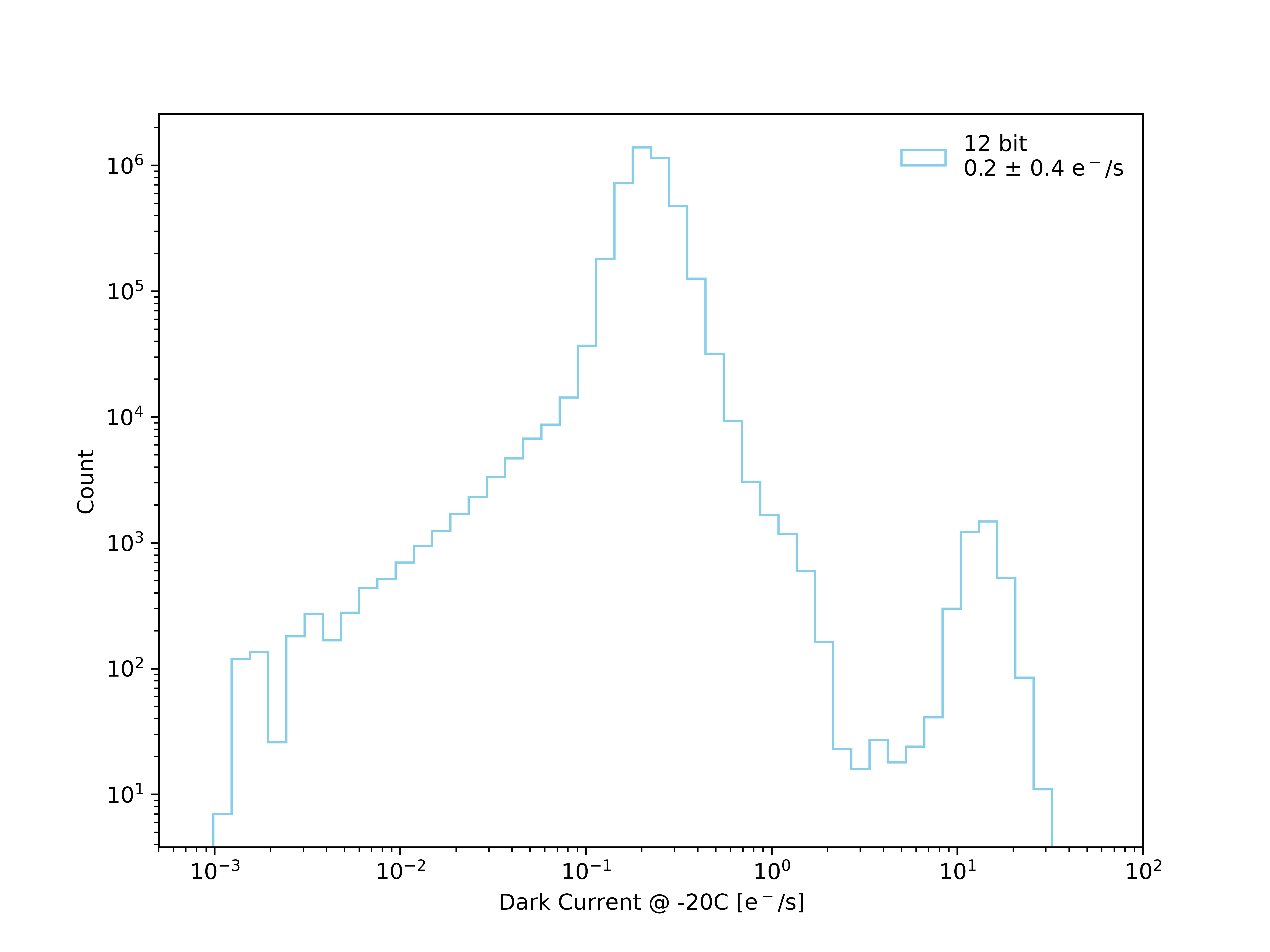}
  \caption{Histogram showing the distribution of each pixel's dark current level in the Sona-6 for the 12-bit mode at $-20^\circ$C.}
  \label{fig: DC_6}
\end{figure}

\begin{figure}[h!]
  \centering
  \includegraphics[width = 0.75\columnwidth]{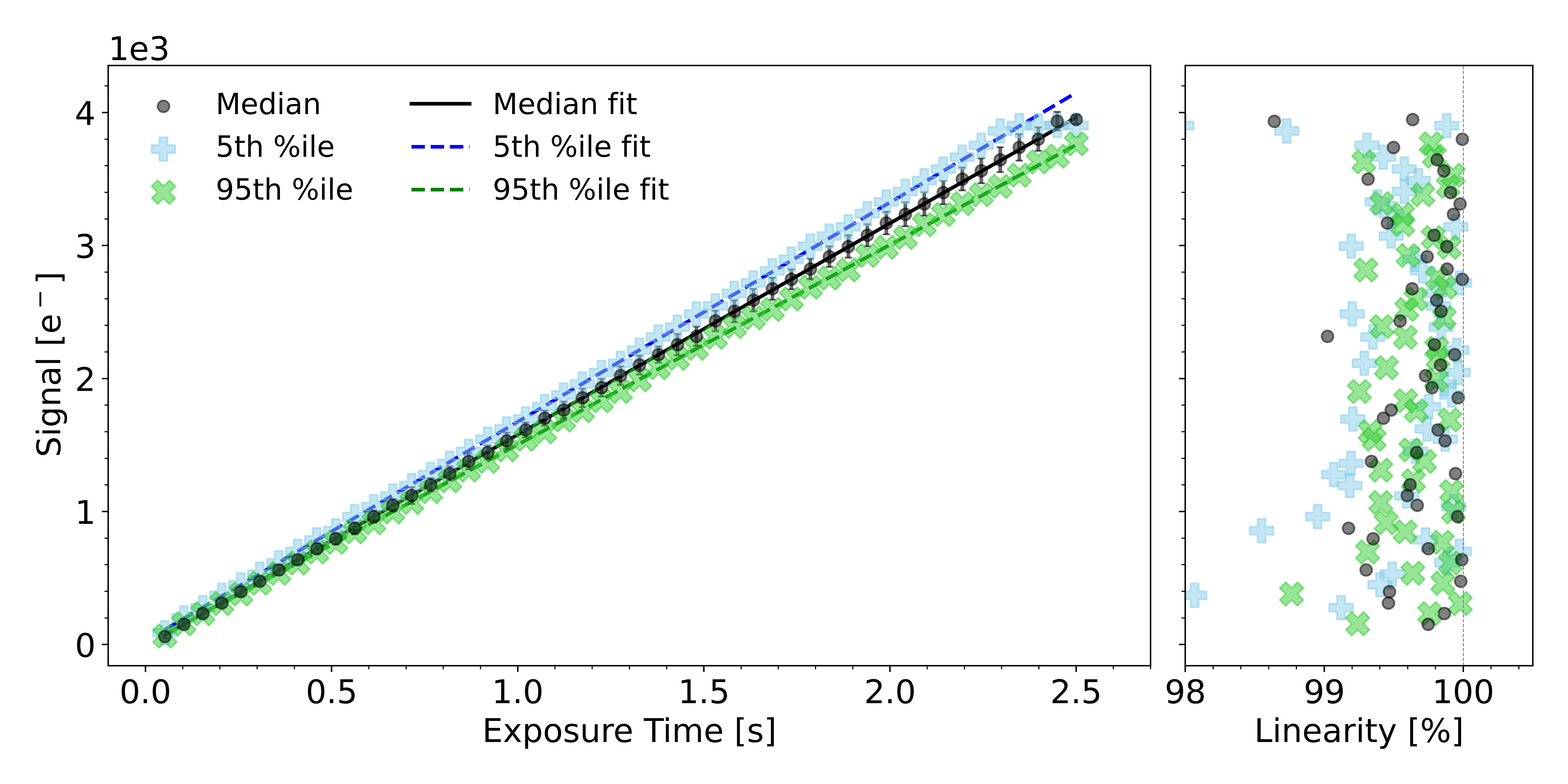}
  \caption{Linearity plots for the Sona-6's 12-bit mode. The plot shows the performance of the pixels with the median gain (black), 5th percentile gain (blue), and 95th percentile gain (green). The residual plot on the right illustrates how close each point is to perfect linearity, expressed as a percentage.}
  \label{fig: lin_6}
\end{figure}

\newpage

\subsection{Ximea xiJ MJ042MR-GP-P6-BSI}    \label{sec: ximea}

\begin{figure}[h!]
  \centering
  \includegraphics[width=\columnwidth]{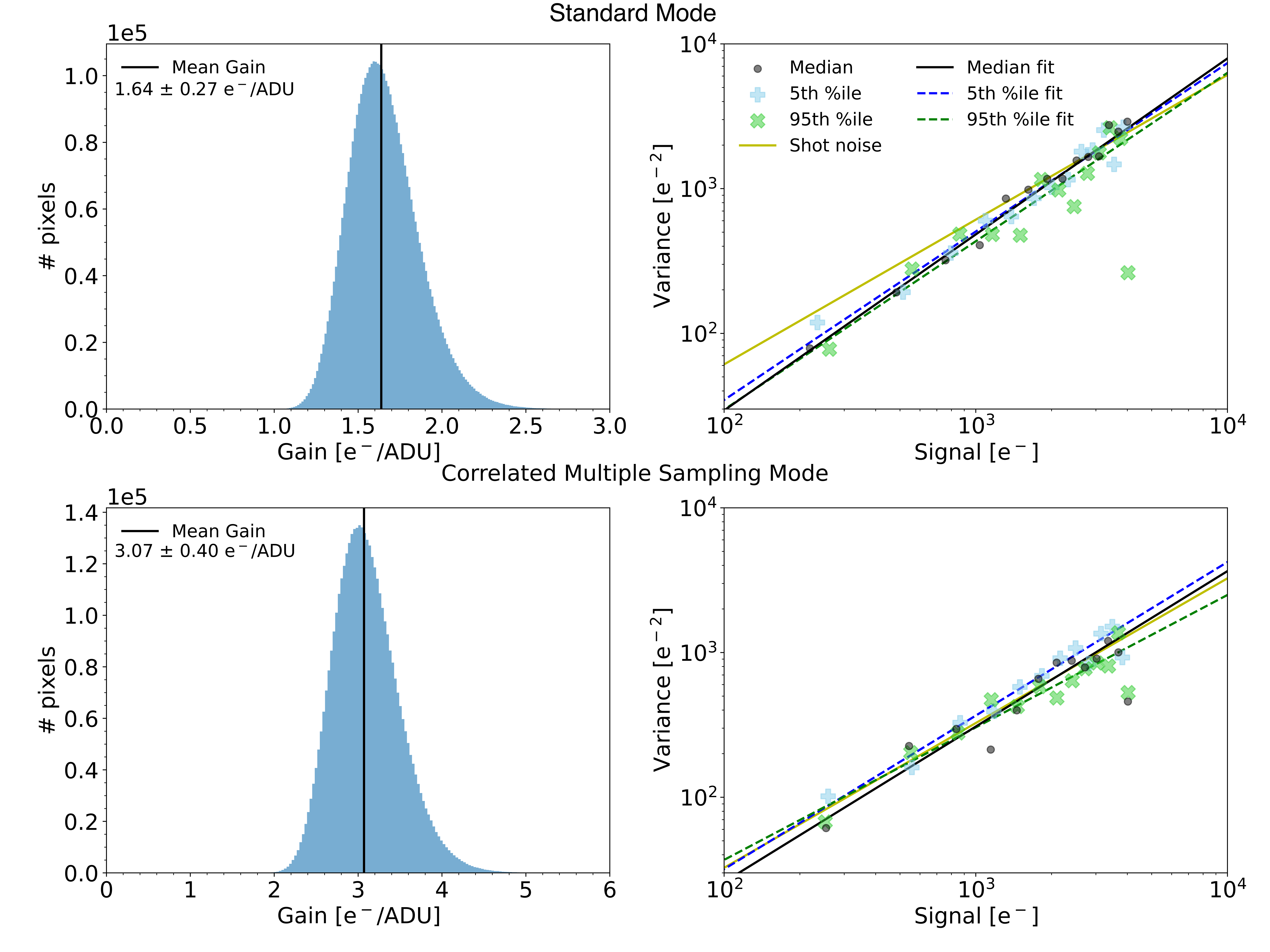}
\caption{Results of the Ximea xiJ's conversion gain calculations for two modes: standard (top) and correlated multiple sampling (bottom). \textit{Left:} Histograms showing the distribution of conversion gain values across all pixels in the Ximea xiJ. The solid black line represents the mean gain across all pixels. \textit{Right:} Mean-variance log plots for pixels with median gain, 5th percentile gain, and 95th percentile gain. The solid yellow line represents the median shot noise. \underline{Note:} Interestingly, the data in the standard mode's mean-variance plot falls below the shot noise curve. This is physically impossible and suggests that the camera exhibits less noise than the statistical Poisson noise in incident photons. Additional data is needed to investigate this behavior further.}
  \label{fig: gain_ximea}
\end{figure}

\begin{figure}[h]
  \centering
  \includegraphics[width=0.65\columnwidth]{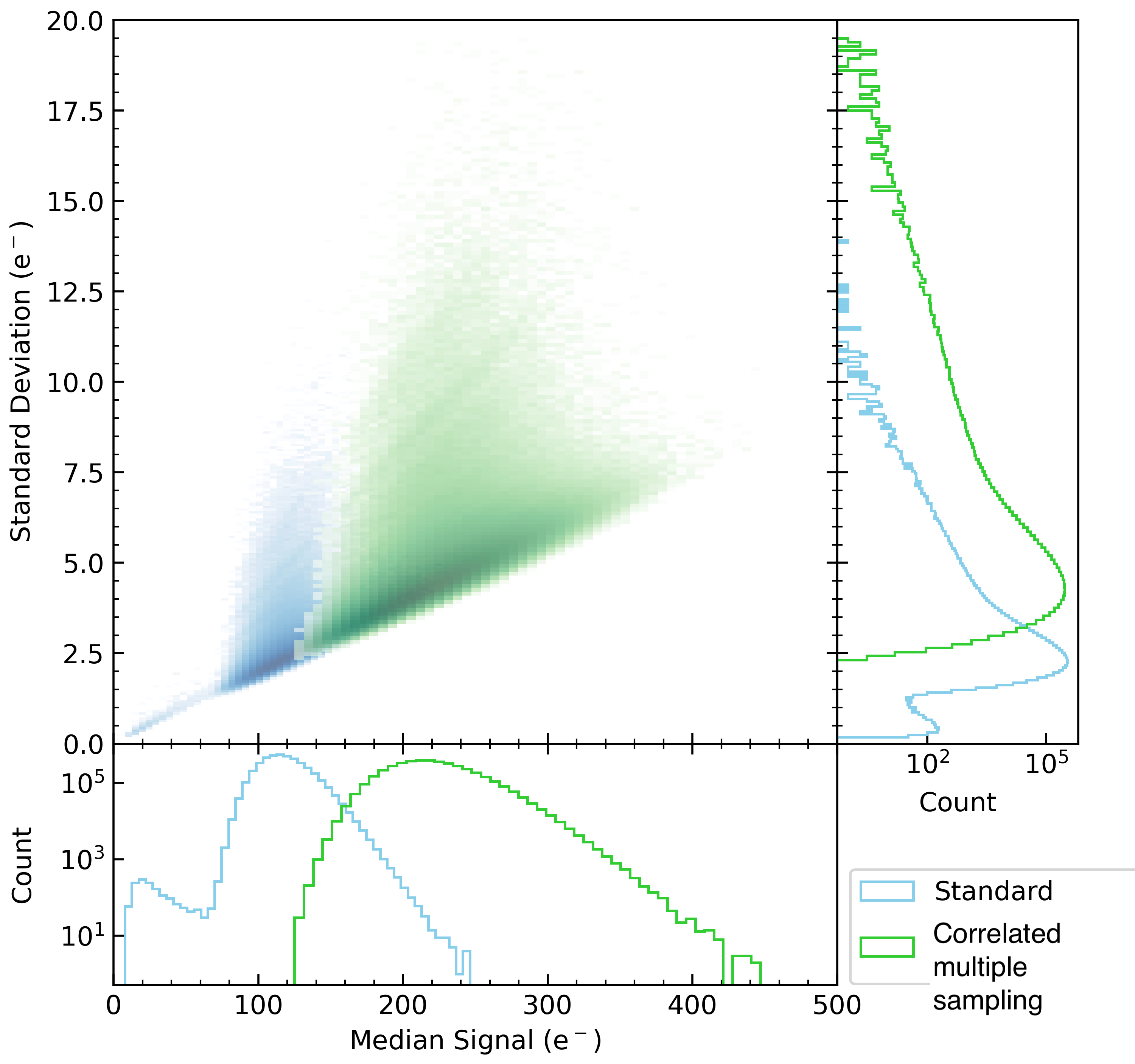}
  \caption{A 2-dimensional histogram showing the distribution of each pixel's read noise (standard deviation) and bias (median signal) levels in the Ximea xiJ for the standard (blue) and correlated multiple sampling (green) modes. The side 1-dimensional histograms display the read noise distribution on the right and the bias distribution on the bottom.}
  \label{fig: RN_ximea}
\end{figure}

\begin{figure}[h]
  \centering
  \includegraphics[width=0.8\columnwidth]{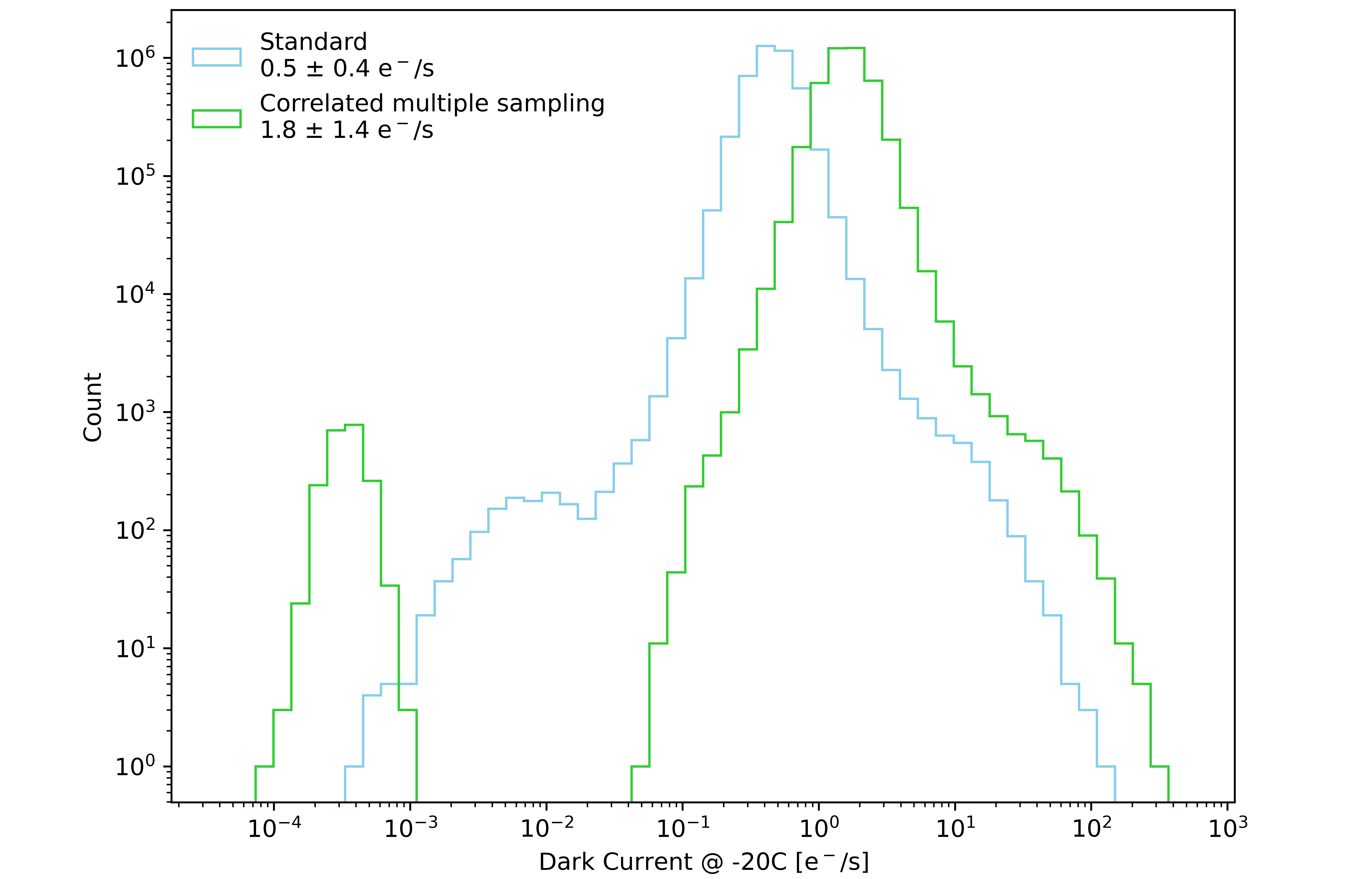}
  \caption{Histograms showing the distribution of each pixel's dark current level in the Ximea xiJ for the standard (blue) and correlated multiple sampling (green) modes at $-20^\circ$C.}
  \label{fig: DC_ximea}
\end{figure}

\begin{figure}[h!]
  \centering
  \includegraphics[width = 0.8\columnwidth]{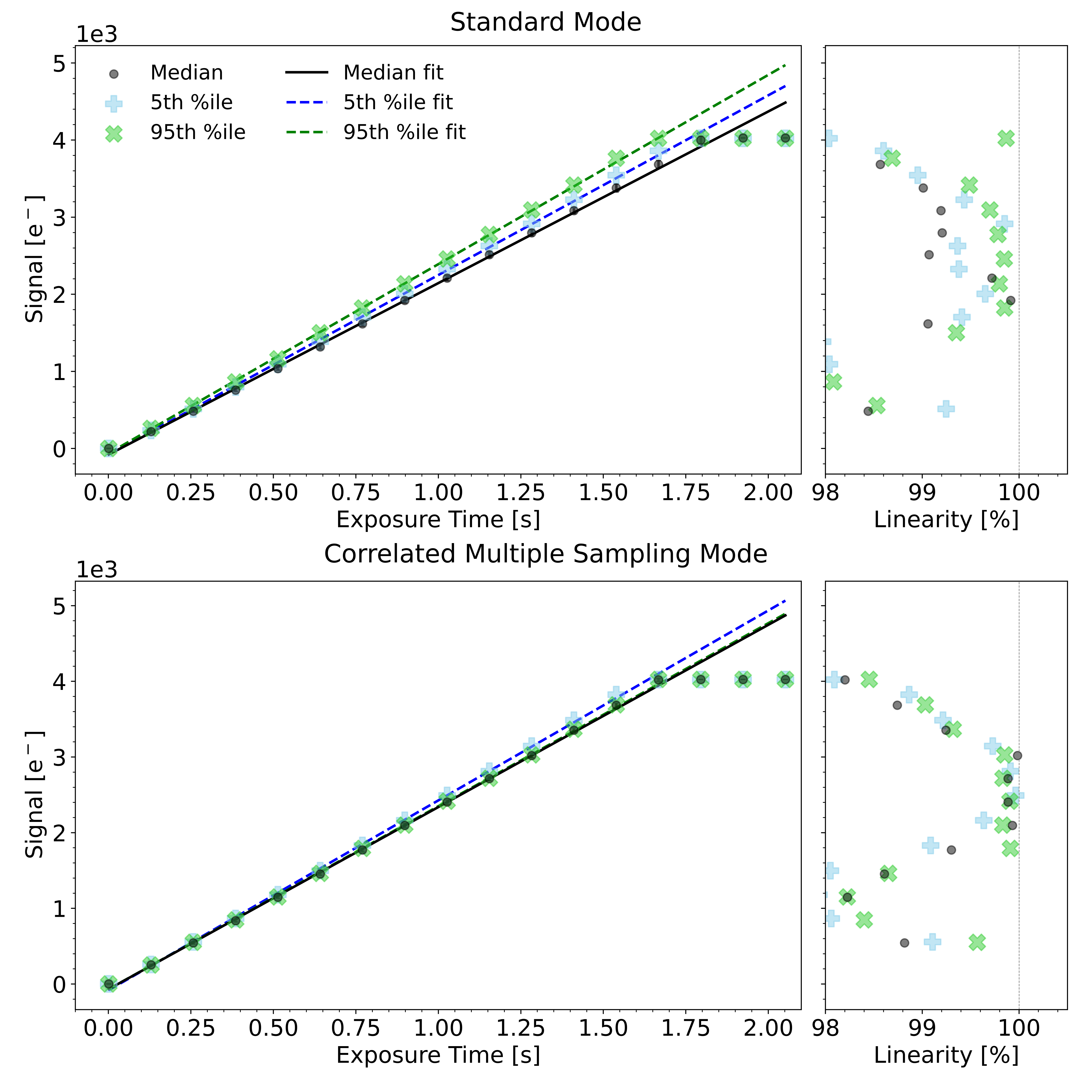}
  \caption{Linearity plots for the Ximea xiJ's two modes: standard (top) and correlated multiple sampling (bottom). Each plot shows the performance of the pixels with the median gain (black), 5th percentile gain (blue), and 95th percentile gain (green). The residual plot on the right illustrates how close each point is to perfect linearity, expressed as a percentage.}
  \label{fig: lin_ximea}
\end{figure}

\end{document}